\documentclass[reprint,aps,prb,superscriptaddress]{revtex4-1}
\usepackage{natbib} 
\usepackage{textcomp} 

\usepackage{bibunits}
\defaultbibliographystyle{apsrev4-2}

\usepackage{graphicx}
\usepackage{dcolumn}
\usepackage{bm}
\usepackage{soul,xcolor}
\usepackage{graphicx}
\usepackage{epstopdf}
\usepackage{color}
\usepackage{float} 
\usepackage{graphics}
\usepackage{bm}
\usepackage{amsmath} 
\usepackage{amsfonts}
\usepackage{amssymb}
\usepackage{makecell}
\usepackage{threeparttable}
\usepackage{lineno}
\usepackage{siunitx}
\usepackage{hyperref}

\soulregister\cite7
\soulregister\ref7
\soulregister\textbf7
\soulregister\textit7
\usepackage{soul,xcolor}
\soulregister\cite7
\soulregister\ref7
\soulregister\textbf7
\soulregister\textit7
\soulregister\textsubscript7
\setcitestyle{super}

\newcommand*{\citen}{}
\DeclareRobustCommand*{\citen}[1]{%
  \begingroup
    \romannumeral-`\x 
    \setcitestyle{numbers}%
    \cite{#1}%
  \endgroup
}

\begin{document}

\begin{bibunit}[apsrev4-2]
\preprint{APS/123-QED}
\title{Observation of room temperature intrinsic nonlinear thermoelectric effects\\ in low-dimensional semimetals}


\author{Kurea Nakagawa$^*$}

\author{Krishnaraajan Sundararajan$^\dag$}

\author{C\'{e}dric A. Cordero-Silis}

\author{Bart J. van Wees}%
\author{Marcos H.D. Guimarães}
\affiliation{Zernike Institute for Advanced Materials, University of Groningen, NL-9747 AG Groningen, The Netherlands}%


 

\begin{abstract}
Nonreciprocal control of thermoelectric responses offers a promising strategy for next-generation thermal-management and energy-conversion. While nonlinear electrical transport has recently emerged as an intrinsic property of low-symmetry quantum materials, their thermoelectric counterparts have not been demonstrated. Here, exploiting harmonic detection with gradient-reversal techniques, we report intrinsic nonlinear thermoelectric responses up to room temperature in the low-symmetry type-II Weyl semimetals $T_\mathrm{d}$-WTe$_2$ and TaIrTe$_4$ in the absence of magnetic fields or magnetic materials. We resolve all symmetry-allowed components of the second-order thermoelectric tensor, including the nonlinear Seebeck, nonlinear Nernst, and nonlinear mixed-directional thermoelectric effects and demonstrate that both Berry-curvature-related and scattering-induced contributions govern the different nonlinear thermoelectric responses. Our results show that nonlinear thermoelectricity arises intrinsically from reduced crystal symmetry and that engineering effects related to scattering in these materials provides a versatile platform for exploring higher-order heat-to-charge current conversion beyond the linear response.
\begin{center}
\footnotesize
*$^\dag$
Corresponding authors. These authors contributed equally to this work. \\
Corresponding authors: *k.nakagawa@rug.nl, $^\dag$ k.sundararajan@rug.nl
\end{center}
\end{abstract}

\maketitle
\makeatletter
\global\let\frontmatter@footnote@produce\@empty
\makeatother

\section{Introduction} 
Harnessing thermal responses in nanoscale electronic devices opens new avenues for functionalities beyond conventional charge-based transport. Well-known thermoelectric phenomena, such as the Seebeck \cite{ashcroft1976solid} and Nernst effects~\cite{Nernst}, generate electrical voltages which are linearly proportional to an applied temperature gradient, either longitudinal or transversal to it, respectively, with the latter requiring the breaking of time-reversal symmetry. These linear responses are fundamentally reciprocal in nature, i.e., their magnitude remains unchanged upon reversing the temperature gradient, and vanish in the absence of macroscopic temperature gradients. On the other hand, nonlinear thermoelectric responses, wherein the resulting electric field does not reverse upon the reversal of the driving temperature gradient, generate rectified electrical outputs under a thermal bias, providing a route to thermal-to-electrical response rectification enabling functional elements for nonreciprocal thermal sensing and energy conversion~\cite{malik2022review}.\\ 

Nonlinear transport phenomena emerging beyond the linear-response regime within single crystalline materials require lowering of crystal symmetries~\cite{suarez2025nonlinear}. A prototypical example is the nonreciprocal magnetoresistance, which occurs only when both inversion and time-reversal symmetries are broken, as required by the Onsager reciprocity relations, typically through an external magnetic field or spontaneous magnetization~\cite{rikken2001electrical, tokura2018nonreciprocal, jo2025anomalous}. More recently, the nonlinear Hall effect, an intrinsic second-order transverse electrical response, has been observed in low-symmetry materials, even in the presence of time-reversal symmetry~\cite{ma2019observation, du2021nonlinear, kumar2021room, jiang2025probing, shvetsov2019nonlinear, dzsaber2021giant, tiwari2021giant, kiswandhi2022observation, tiwari2021giant}. \\

Experimental realizations of nonlinear thermoelectric effects have mostly relied on externally engineered structural asymmetry through device architectures, such as semiconductor quantum dots~\cite{svensson2013nonlinear, svilans2016nonlinear}, molecular and nanojunction systems~\cite{reddy2007thermoelectricity, prakash2022thermoelectric}, and heterostructure devices~\cite{arisawa2024observation,hirata2025nonlinear}, which depend critically on structural design and fabrication complexity, limiting their generality and scalability. A more robust approach is offered by low-symmetry materials, wherein nonlinear thermoelectric effects arise intrinsically from microscopic properties governed by the symmetry constraints of the crystal. In this context, the nonlinear Nernst (NLN) effect in the absence of an external magnetic field has been theoretically proposed to share a common symmetry origin with the nonlinear Hall effect, and has been mainly attributed to the Berry curvature dipole~\cite{nakai2019nonreciprocal, zeng2019nonlinear, zeng2020fundamental, varshney2025intrinsic, zeng2022chiral,yu2019topological, sodemann2015quantum} or scattering mechanisms~\cite{varshney2026asymmetric,nakai2019nonreciprocal}.
Despite considerable theoretical interest, experimental demonstrations of intrinsic nonlinear thermoelectric effects remain scarce, in contrast to the growing number of observed nonlinear electrical responses~\cite{
kang2019nonlinear, huang2023intrinsic, wang2023quantum, cheng2024giant}. Very recently, the observation of the NLN effect in trilayer graphene has been suggested~\cite{liu2025nonlinear}. Nonetheless, the effect was only detected at ultra-low temperatures, and the separation between temperature-dependent linear and nonlinear contributions was not addressed. As we show, it is crucial to separate the intrinsic nonlinear thermoelectric effects from the temperature dependence of the linear effects as it can affect or even completely dominate the measured higher order responses.\\ 

\begin{figure*}[t]
   \centering
   \includegraphics[width=\textwidth]{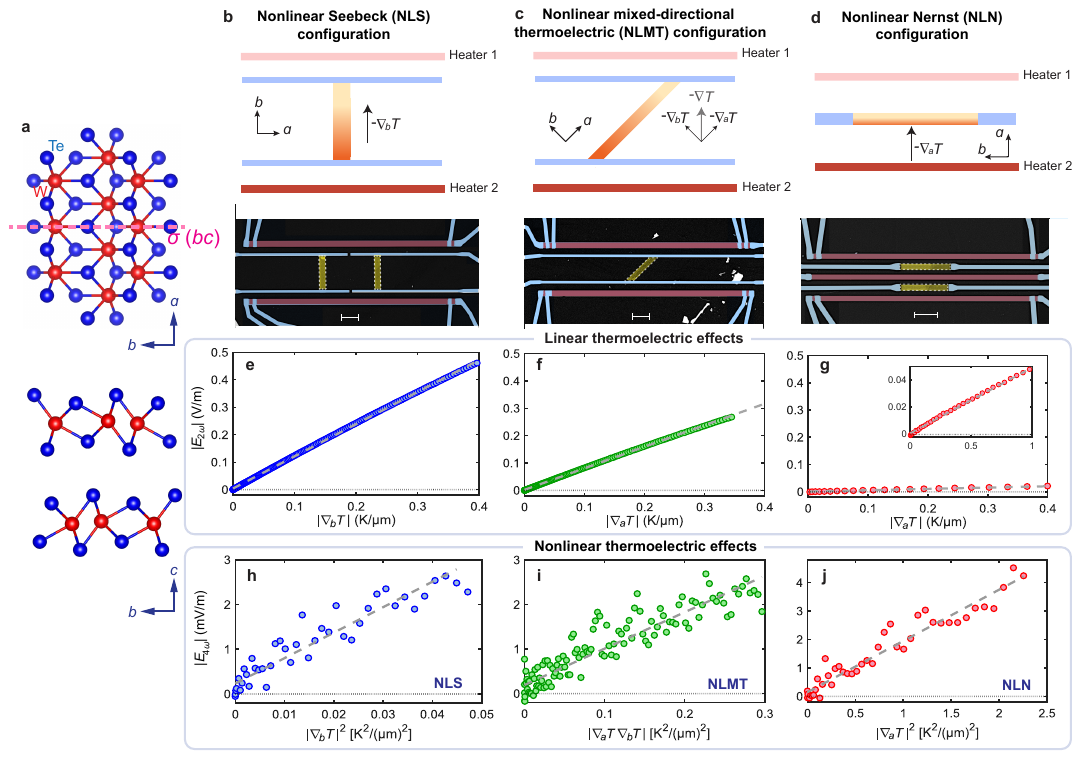}   
   \caption{\textbf{Linear/nonlinear thermoelectric responses in WTe\textsubscript{2}} 
   \textbf{a}, Crystal structure of $T_\mathrm{d}$-WTe\textsubscript{2}, with the mirror plane [$\sigma$($bc$)] indicated by the pink dashed line. \textbf{b}-\textbf{d},
   Schematic illustrations (top) and corresponding false-coloured scanning electron microscopy (SEM) images (bottom) of the devices designed to measure the NLS (\textbf{b}), NLMT (\textbf{c}), and NLN (\textbf{d}) effects. In the SEM images, WTe\textsubscript{2}, Pt heaters, and Au contacts are shown in yellow, red, and light blue, respectively (Scale bar: $\SI{10} {\micro\meter}$). 
   \textbf{e}–\textbf{g}, Second-harmonic thermoelectric responses at $T =$ 35 K, plotted as a function of the applied temperature gradient for the three devices shown in \textbf{b}–\textbf{d}, respectively. The axis scales are unified across all panels, with an inset showing a magnified plot for panel~\textbf{g}.
   \textbf{h}–\textbf{j}, Fourth-harmonic thermoelectric responses at $T =$ 35 K, plotted as a function of the square of the applied temperature gradient for the same devices. The data are symmetrized with respect to temperature-gradient reversal to isolate intrinsic nonlinear thermoelectric contributions (see text). Dashed lines indicate linear fits.}
   \label{fig1}
\end{figure*}

Here we unambiguously demonstrate intrinsic nonlinear thermoelectric effects arising from low crystal symmetry that persist even up to room temperature. Utilizing the low-symmetry type-II Weyl semimetals WTe\textsubscript{2} and TaIrTe\textsubscript{4}, we observe all the symmetry allowed components of the second-order thermoelectric tensor. Our results include the first observation of the nonlinear Seebeck, the nonlinear Nernst, and a nonlinear mixed-directional thermoelectric effect in a single material. We further find that scattering effects play a crucial role in the observed nonlinear thermoelectric responses. This further establishes intrinsic nonlinear thermoelectric transport as a symmetry-governed phenomenon that does not rely on engineered structural asymmetry or time-reversal symmetry breaking. 


\section{Results and discussion}

\subsection*{Symmetry-allowed nonlinear thermoelectric responses in thin WTe\textsubscript{2} and TaIrTe\textsubscript{4}}



A material's thermoelectric response is governed by its crystal symmetry, which restricts the electric-field response to an applied temperature gradient. Since both the temperature gradient $\nabla T$ and the induced electric field $E$ are polar vectors, they transform identically under the symmetry operations of the crystal. In the absence of an external magnetic field, the linear thermoelectric response is described by a second-rank tensor, while the nonlinear thermoelectric effects by a third-rank tensor that couples quadratically to the temperature gradient. Crucially, the transformation properties of these tensors under the crystal symmetries determine which responses are forbidden.\\

\begin{figure*}[t]
   \centering
   \includegraphics[width=\textwidth]{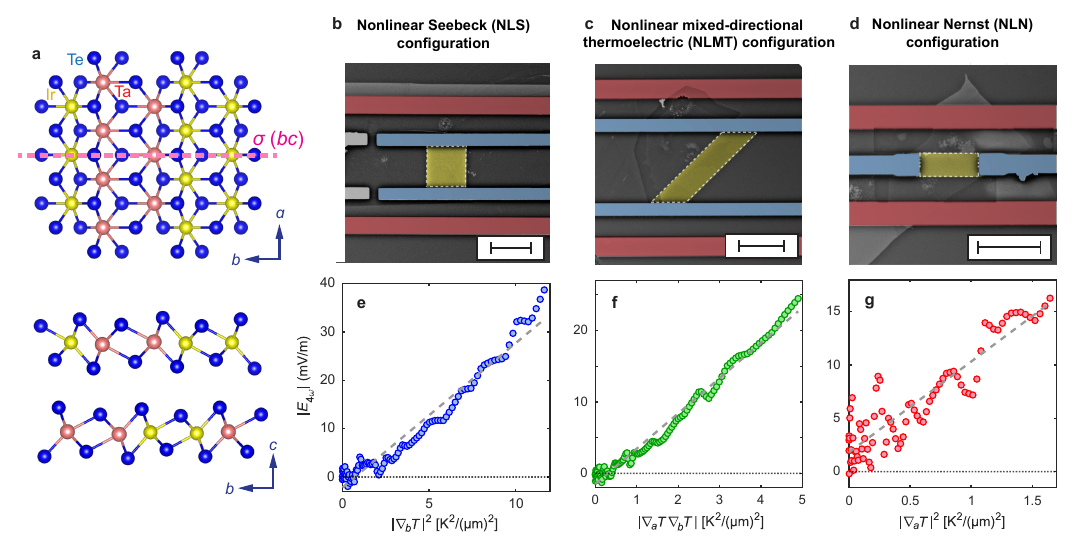}   
   \caption{\textbf{Nonlinear thermoelectric responses in TaIrTe\textsubscript{4} at room temperature} 
   \textbf{a}, Crystal structure of TaIrTe\textsubscript{4}, with the mirror plane [$\sigma$($bc$)] indicated by the pink dashed line. \textbf{b}-\textbf{d},
   False-coloured SEM images of the devices designed to measure the NLS (\textbf{b}), NLMT (\textbf{c}), and NLN (\textbf{d}) effects. In the SEM images, TaIrTe\textsubscript{4}, Pt heaters, and Au contacts are shown in yellow, red, and light blue, respectively (Scale bar: $\SI{5} {\micro\meter}$). 
   \textbf{e}–\textbf{g}, Fourth-harmonic thermoelectric responses at room temperature ($T =$ 292.5 K), plotted as a function of the square of the applied temperature gradient for the same devices. The data are symmetrized with respect to temperature-gradient reversal to isolate intrinsic nonlinear thermoelectric contributions (see text). Dashed lines indicate linear fits.}
   \label{fig2}
\end{figure*}

Exfoliated WTe$_2$ and TaIrTe$_4$, owing to their low crystal symmetry, provide an ideal platform with minimal symmetry constraints for exploring nonlinear thermoelectric responses. Thin $T_\mathrm{d}$-WTe$_2$ and TaIrTe$_4$ flakes, both pertaining to space group $Pm$ \cite{WTe_Torque,TIT_Torque}, retain only a single mirror symmetry with respect to the $bc$ plane [$\sigma(bc)$], whereas the screw-axis and glide-plane symmetries present in the bulk crystal (space group $Pmn2_1$) are broken (Figs.~\ref{fig1}\bf{a} \normalfont and \ref{fig2}\bf{a}\normalfont). In the presence of the mirror symmetry, the linear and nonlinear thermoelectric responses in the two-dimensional $a$–$b$ crystal-plane can be expressed as: 

\begin{equation}
\label{Reduced_Matrix_form_linear}
    \left[\begin{array}{c}E_a \\ E_b\end{array}\right]
    =
    \left[\begin{array}{ccc}
        S_{aa}^{(1)}& 0 \\
        0 & S_{bb}^{(1)}
    \end{array}\right]
    \left[\begin{array}{c}
        (-\nabla_a T) \\
        (-\nabla_b T)
    \end{array}\right],
\end{equation}

\begin{equation}
\label{Reduced_Matrix_form_nonlinear}
    \left[\begin{array}{c}E_a \\ E_b\end{array}\right]
    =
    \left[\begin{array}{ccc}
        0 & S_{aab}^\mathrm{(2)} & 0 \\
        S_{baa}^\mathrm{(2)} & 0 & S_{bbb}^\mathrm{(2)}
    \end{array}\right]
    \left[\begin{array}{c}
        (-\nabla_a T)^2 \\
        2(-\nabla_a T)(-\nabla_b T) \\
        (-\nabla_b T)^2
    \end{array}\right],
\end{equation}
where $S_{ij}^{(1)}$ and $S^\mathrm{(2)}_{ijk}$ are the linear and second-order thermoelectric response tensors, where the indices $a$ and $b$ denote directions along the crystalline axes (Supplementary Section I).\\

To verify these symmetry-allowed linear and nonlinear thermoelectric effects, we fabricate devices with three distinct geometries (Figs.~\ref{fig1}\bf{b}\normalfont-\bf{d}\normalfont). To minimize parasitic temperature gradients, the heaters were made more than five times longer than the voltage detection channel, and the contact electrodes' geometry is kept symmetric relative to the heaters. The magnitude of the temperature gradient was derived from finite element simulations (Supplementary Section XII).\\

We observe linear thermoelectric responses in all three devices, as shown in Figs.~\ref{fig1}\textbf{e}–\textbf{j} for WTe$_2$.
According to the symmetry constraints given in Eq.~(\ref{Reduced_Matrix_form_linear}), the mirror symmetry forbids the generation of a transverse thermopower, allowing only the longitudinal response, i.e., the Seebeck effect. The linear Seebeck effect accounts for the responses observed in Figs.~\ref{fig1}\textbf{e} and \textbf{f}, which correspond to $S_{bb}^{(1)}$ and $S_{aa}^{(1)}$, respectively (Supplementary Section I). The linear transverse response although expected to vanish in the absence of an external magnetic field, in our devices as shown in Fig.~\ref{fig1}\textbf{j}, exhibits a response that is approximately one order of magnitude smaller than those of the other two configurations. We attribute this residual response to the transverse Seebeck effect, which results from a slight misalignment of the device channel with respect to the crystal axes, combined with strong thermoelectric anisotropy in the Seebeck effect (Supplementary Section V).\\

As the heater power depends quadratically on the applied heater current, linear thermoelectric responses are detected in the second-harmonic voltage response, while nonlinear thermoelectric responses appear in the fourth-harmonic voltage response. In order to properly isolate the intrinsic nonlinear thermoelectric responses, it is critical to take into consideration the temperature dependence of the linear thermoelectric effects $S_{ij}^\mathrm{(1)}$. The total fourth-harmonic response ($E_{i, 4\omega}$) is given by
\begin{equation}
\label{Eq:4omega}
    E_{i,4\omega} = -\left(\left.\frac{\partial S_{ij}^\mathrm{(1)}}{\partial T}\right|_{T_0} \cdot \Delta T\right) |\nabla_j T| + S_{ijk}^\mathrm{(2)} (\nabla_j T)(\nabla_k T),
\end{equation}
where the second term on the right side is the intrinsic nonlinear thermoelectric effect and the first term arises from the temperature dependence of the linear thermoelectric effect. Here, exploiting the directional nature of linear thermoelectric effects, we effectively isolate the intrinsic nonlinear response from the temperature-dependent linear response through symmetrization of the responses by alternating the active heater electrode (Supplementary Section VI). We additionally rule out contributions to the measured response arising from the temperature dependence of the heater resistance (Supplementary Section VI). The isolated intrinsic nonlinear signals are plotted in Figs.~\ref{fig1}\bf{h}\normalfont-\bf{j}\normalfont. An equivalent approach to isolating these signals was reported in Ref.~\citen{arisawa2024observation}, wherein the isolation is achieved by introducing a relative phase difference between the two heater currents, while in contrast, our method employs using two heater electrodes without a relative phase shift.
\\

We observe a clear nonlinear thermoelectric response in all three geometries for WTe$_2$, though they exhibit markedly different magnitudes across the various device geometries (note the different \textit{x}-axis scales for Figs.~\ref{fig1}\bf{h}\normalfont-\bf{j}\normalfont).
These three nonlinear thermoelectric responses correspond to three of the six independent second-order thermoelectric tensor components in Eq.~(\ref{Reduced_Matrix_form_nonlinear}) allowed by the crystal symmetry: the nonlinear Seebeck (NLS) effect [$S_{bbb}^\mathrm{(2)}$], wherein the induced voltage is collinear to the temperature gradient~(Fig.~\ref{fig1}\bf{h}\normalfont); the NLN effect [$S_{baa}^\mathrm{(2)}$], wherein the voltage is generated transverse to the temperature gradient under zero magnetic field~(Fig.~\ref{fig1}\bf{i}\normalfont); and a nonlinear mixed-directional thermoelectric (NLMT) effect [$S_{aab}^\mathrm{(2)}$], wherein the generated voltage depends on the temperature gradients along two orthogonal crystallographic directions and is along the crystallographic \textit{a}-axis~(Fig.~\ref{fig1}\bf{j}\normalfont). To verify the role of crystal symmetry, we fabricated an additional device designed to probe the symmetry-forbidden components. Within our experimental resolution, we observe no detectable nonlinear thermoelectric response (Supplementary Section VII), confirming that the nonlinear thermoelectric effects reported here are consistent with the symmetry constraints.\\

Motivated by the observation of a large nonlinear Hall effect at room temperature in TaIrTe$_4$~\cite{kumar2021room,jiang2025probing}, we investigate its nonlinear thermoelectric responses (Fig.~\ref{fig2}). Since TaIrTe$_4$ and WTe$_2$ share the same crystal symmetry, the symmetry-allowed components of the nonlinear thermoelectric response in the two-dimensional $a$–$b$ plane are expected to follow Eq.~(\ref{Reduced_Matrix_form_nonlinear}). Using the measurement scheme and the device geometries introduced earlier (Fig.~\ref{fig2}\textbf{b}–\textbf{d}), we observe pronounced nonlinear thermoelectric responses at zero magnetic field in all three geometries, corresponding to the NLS, NLN and NLMT effects (Fig.~\ref{fig2}\textbf{e}–\textbf{g}) up to room temperature. We note that when the heater current is increased further, the responses deviate from a linear fit, possibly due to contributions from higher-order responses.

\subsection*{Temperature dependence of the nonlinear thermoelectric effects in WTe\textsubscript{2} and TaIrTe\textsubscript{4}}
\begin{figure*}[htb]
   \centering
   \includegraphics[width=\textwidth]{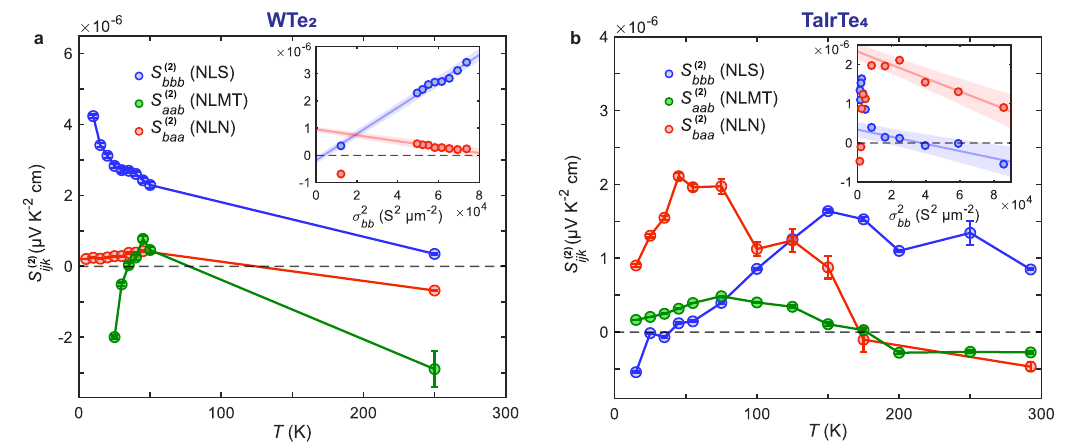}   
   \caption{\textbf{Nonlinear thermoelectric coefficients in WTe\textsubscript{2} and TaIrTe\textsubscript{4}}
   \textbf{a}-\textbf{b}, Temperature dependence of the nonlinear thermoelectric coefficients: NLS [$S_{bbb}^\mathrm{(2)}$], NLN [$S_{baa}^\mathrm{(2)}$], NLMT [$S_{aab}^\mathrm{(2)}$] in WTe\textsubscript{2} (\textbf{a}) and TaIrTe\textsubscript{4} (\textbf{b}). The insets show the NLS and NLN coefficients as functions of the square of the electrical conductivity along $b$-axis; $\sigma_{bb}$. Linear fits to the data below 55 K are shown, with the shaded regions representing the standard deviation of the fits.}
   \label{fig3}
\end{figure*}

The temperature dependence of the nonlinear thermoelectric coefficients in WTe\textsubscript{2} and TaIrTe\textsubscript{4} are shown in Fig.~\ref{fig3}. We observe that each of the symmetry-allowed nonlinear thermoelectric effect exhibits a characteristic temperature dependence distinct from each other. For WTe\textsubscript{2} (Fig.~\ref{fig3}a), the NLS coefficient exhibits a maximum at the lowest temperature of 15 K and decreases monotonically with increasing temperature, exhibiting a behavior distinct from that of the linear Seebeck effect. In contrast, the NLN and NLMT coefficients display a local maximum around 50 K, and show a sign change between 50 K and 250 K. As an overall trend, the NLS and NLMT effects exhibit a larger response in comparison to that of the NLN effect.\\

For TaIrTe\textsubscript{4} (Fig.~\ref{fig3}b), the NLS coefficient increases gradually from low temperatures, reaches a maximum at around 150 K, and decreases slightly while remaining relatively large up to room temperature. The NLN and NLMT coefficients both show a local maximum around 50 to 75 K and change sign near 175 K. The temperatures at which the NLS and NLN coefficients reach their maxima are close to those where the Seebeck and Nernst coefficients exhibit local maxima (Supplementary Section IV). Remarkably, in both materials, the nonlinear thermoelectric coefficients near room temperature remain on the same order of magnitude as those at low temperatures, and in some cases are even larger. We note that the sign of the nonlinear thermoelectric coefficients should be interpreted with caution, as it changes under reversal of the crystal axes. Although the crystal orientation can be determined experimentally, the crystal axis direction cannot be uniquely specified in the present measurements, preventing an unambiguous determination of the sign.\\

Our measurements demonstrate the robustness of nonlinear thermoelectric effects at elevated temperatures. We observe sizable nonlinear responses persist up to room temperature, indicating an origin in intrinsic band-structure and scattering mechanisms that persist despite strong phonon scattering. The NLS coefficient reaches a maximum value of $\SI{4.3e-6}{\micro V\,K^{-2}\,cm}$ in WTe$_2$ and has a room temperature value of $\SI{0.9e-6}{\micro V\,K^{-2}\,cm}$ in TaIrTe$_4$. To our knowledge, this represents the first observation of three distinct nonlinear thermoelectric effects within a single material driven by the lowered crystal symmetry over such a broad temperature range, with a magnitude exceeding that reported in magnetic heterostructures~\cite{hirata2025nonlinear}.

To elucidate the microscopic origin of the nonlinear thermoelectric responses, we investigate the temperature dependence of the nonlinear thermoelectric coefficients as a function of the corresponding longitudinal conductivity. As shown in the insets of Fig.~\ref{fig3}, we employ the scaling relation $S_{ijj}^{(2)} = A\sigma_{ii}^{2} + C$, where $\sigma_{ii}$ denotes the longitudinal electrical conductivity. Within the framework of semiclassical Boltzmann transport theory, $\sigma_{ii}$ is proportional to the scattering time ($\tau$) and therefore, the slope ($A$) and intercept ($C$) correspond to contributions from skew scattering and from Berry curvature dipole, respectively~\cite{varshney2026asymmetric}. The validity of the present scaling analysis, as well as comparisons with alternative scaling relations, are discussed in Supplementary Section XI.\\

Below 55 K, where the Mott–Boltzmann framework remains valid, both the NLN and NLS responses in WTe\textsubscript{2} and TaIrTe\textsubscript{4} are well described by the scaling relation $S_{ijj}^{(2)} = A\sigma_{ii}^{2} + C$. At higher temperatures, phonon scattering and the temperature dependent Fermi level shift complicate the picture, hence we focus on the low temperature behaviour.
While both the NLS and NLN coefficients exhibit a clear $\sigma_{ii}^{2}$ dependence, the NLS response shows a negligible offset, whereas the NLN response exhibits a finite offset.
These results indicate that the NLS response is dominated by skew-scattering contributions, while the NLN response contains both Berry curvature dipole and skew-scattering mechanisms. This interpretation is further supported by recent theoretical work that indicate that Berry-curvature-dipole-induced nonlinear thermoelectric responses cannot contribute to the NLS effect \cite{varshney2026asymmetric}.
The NLN effect seems to share a common microscopic origin with the nonlinear Hall effect (Supplementary Section I), where the Berry curvature dipole is claimed to be the main mechanism~\cite{zeng2019nonlinear,zeng2020fundamental, ma2019observation, kumar2021room, jiang2025probing} with skew-scattering and side-jump processes potentially also contributing to the observed responses~\cite{kang2019nonlinear,nakai2019nonreciprocal}.\\ 


\section{Conclusions}
The experimental observation of nonlinear thermoelectric transport in exfoliated WTe\textsubscript{2} and TaIrTe\textsubscript{4} establishes a direct link between reduced crystal symmetry and higher-order heat current to charge current interconversion in low dimensional quantum materials. This highlights a robust route toward directional thermal control and energy conversion based on intrinsic material properties, and potentially as well as through engineering scattering processes, rather than relying on complex device architectures. 
Extending this approach to other low-symmetry and topological materials in gated hetero-structures, as well as developing microscopic theoretical descriptions of nonlinear thermoelectric coefficients, will be essential steps toward exploiting nonlinear thermoelectricity for next-generation thermal management and solid-state energy technologies.


\section*{Acknowledgements}

The authors acknowledge the technical support from J. G. Holstein, H. H. de Vries, F. H. van der Velde, H. Adema and A. Joshua. The authors acknowledge the research program “Materials for the Quantum Age” (QuMat) for financial support. This program (registration number 024.005.006) is part of the Gravitation program financed by the Dutch Ministry of Education, Culture and Science (OCW). This work was additionally supported by the Zernike Institute for Advanced Materials (ZIAM) at University of Groningen. The device fabrication and characterization were performed using NanolabNL facilities. This work was further supported by the Dutch Research Council (NWO) through grant OCENW.XL21.XL21.058, the Spinoza prize awarded to B.J. van Wees in 2016, as well as by the European Research Council (ERC) under the European Union’s 2DMAGSPIN (Grant No. 101053054) and 2D-OPTOSPIN (Grant No. 101076932). K. Nakagawa is supported by Overseas Research Fellowships of Japan Society for the Promotion of Science. Views and opinions expressed are however those of the author(s) only and do not necessarily reflect those of the European Union or the European Research Council. Neither the European Union nor the granting authority can be held responsible for them.


\section*{Methods}
\subsection*{Device fabrication}
WTe$_2$ flakes were exfoliated from $T_\mathrm{d}$-WTe$_2$ bulk crystals (HQ Graphene) onto 285nm SiO$_2$/Si substrates using the standard scotch-tape method. The flakes with their thickness around 10-30 nm were chosen by optical contrast and transferred onto fused silica substrates by a dry transfer technique. The exact thickness of each flakes was confirmed by atomic force microscopy (Supplementary Section III).
The exfoliation and transfer were performed in a nitrogen-filled glove box with O$_2$ $<$ 0.1 ppm and H$_2$O $<$ 0.5 ppm. The WTe$_2$ flakes were subsequently patterned into rectangular geometries by reactive ion etching using a CF$_4$/O$_2$ plasma, with a poly(methyl methacrylate) (PMMA) mask defined by electron-beam lithography. 

The crystal orientation was independently confirmed by polarized Raman spectroscopy measured on the same devices, as described in the Supplementary Section III. Platinum (Pt) heater strips with a thickness of approximately 40 nm were fabricated adjacent to the WTe$_2$ flakes by electron-beam lithography followed by DC sputtering. Subsequently, Ti/Au electrical contacts (75 nm/5 nm) were defined by electron-beam lithography followed by DC sputtering and electron beam evaporation. Prior to the deposition of the Ti/Au contacts onto both the WTe$_2$ flakes and the Pt strips, the WTe$_2$ surface was cleaned in-situ by argon ion milling to remove any residual native oxide. The final devices were spin-coated with a PMMA layer to protect the WTe$_2$ flakes from further oxidation.\\

The exfoliation and the transfer of TaIrTe$_4$ was done following exactly the same procedure as of WTe$_2$ using bulk crystals from HQ Graphene. Owing to the slow etch rate of TaIrTe$_4$ under CF$_4$/O$_2$ plasma that results in the cross-linking of the PMMA etch mask, the order of contacts and etching for TaIrTe$_4$ was swapped in comparison to that described for WTe$_2$. The etching of the TaIrTe$_4$ flakes was carried out after the fabrication of the heaters and electrical contacts. Aside from this difference, all processes followed the same procedure as that used for WTe$_2$.

In addition, to detect the nonlinear Hall effect and to systematically investigate the anisotropy of the electrical and thermoelectric responses, as well as the distribution of the temperature gradient within the device, we fabricated multiple WTe$_2$ devices with various geometries on Si/SiO$_2$ substrates. The fabrication procedures were identical to those used for the WTe$_2$ devices on fused silica, except for the geometries of the heaters and the samples. The details of the device geometries and the corresponding results are presented in the Supplementary Sections V and X.

\subsection*{Electrical Transport Measurements}
The electrical transport measurements were performed using standard lock-in technique, with an ac current sourced through to the platinum heater at 17.777 Hz (unless otherwise mentioned). 
The voltage response is composed of different orders and can be expressed as
\begin{equation}
    V(t)=\sum_{i=1}^{\infty} R_i I^i(t)
\end{equation}
where $R_i$ is the $i$th-order response of the system to the applied ac current $I(t)$. 
The different harmonic responses of the sample to the driving ac current are measured using lock-in amplifiers (SR830 and AMETEK 7270). The linear thermoelectric effects are detected in the second-harmonic response, and the nonlinear thermoelectric effects are detected in the fourth-harmonic response (Supplementary Section II). We also examined the 1, 3, 5, and 6\textsuperscript{th}-harmonic responses to identify the possible role of capacitive coupling and to rule out the possibility that a parasitic dc current offset causes a mixing of higher order responses into the measured fourth-harmonic response (Supplementary Sections VIII and IX). The frequency dependence of the 2 and 4\textsuperscript{th}-harmonic responses was verified (Supplementary Sections VIII and IX), to ensure that $f < 1/\tau_{\mathrm{th}}$, where $\tau_{\mathrm{th}}$ is the thermal time constant of the substrate.



%

\end{bibunit}

\clearpage
\onecolumngrid

\setcounter{section}{0}
\setcounter{figure}{0}
\setcounter{table}{0}
\setcounter{equation}{0}


\renewcommand{\thefigure}{S\arabic{figure}}
\renewcommand{\thetable}{S\arabic{table}}

\def\cmunits{\text{cm}\textsuperscript{-1}\text{ }}

\begin{center}
{\bfseries\large Supporting Information to ``Observation of room temperature intrinsic nonlinear thermoelectric effects in low-dimensional semimetals''}
\end{center}
\vspace{0.5em}

\section{Effect of crystal symmetry and off-axis temperature gradients on the linear and nonlinear thermoelectric responses}
\label{Sec:Symm}

\subsection{Phenomenological symmetry analysis of nonlinear thermoelectric effects in  WTe\textsubscript{2} and TaIrTe\textsubscript{4}}

The general second-order thermoelectric responses can be written as: 

\begin{equation}
\label{General_NL}
E_i = S_{ijk}^{(2)} (-\nabla_j T)(-\nabla_k T),
\end{equation}
where the response matrix ($\mathbf{S^{(2)}}$) describing the interconversion of an applied temperature gradient into an electric field is a tensor of rank 3 and the indices $i$, $j$ and $k$ denote the directions of the electric field and the temperature gradients, respectively. Although the WTe\textsubscript{2} and TaIrTe\textsubscript{4} studied here are not intrinsically two-dimensional, we restrict our analysis to the in-plane responses, since the measured voltage is confined to two dimensions and since the temperature gradient generated along the in-plane directions are much larger than that oriented out-of-plane (see Sec. \ref{Sec:Grad_OOP} for an estimation and a discussion on the influence of an out-of-plane temperature gradient). Under this restriction, the response tensor reduces to eight components corresponding to $i$, $j$, $k$ = $a$ or $b$, where \textit{a} and \textit{b} denote the principal crystallographic axes of WTe\textsubscript{2} and TaIrTe\textsubscript{4}, that are orthogonal to each other. Noting that of the 8 response elements only 6 are independent (since $S_{aab}^{(2)} = S_{aba}^{(2)}$ and $S_{bab}^{(2)} = S_{bba}^{(2)}$), the generalized non-linear thermoelectric responses as expressed in Eq. \ref{General_NL}, for the case of two-dimensions can be recast into the matrix form: 

\begin{equation}
\label{Matrix_form}
    \left[\begin{array}{c}E_a \\ E_b\end{array}\right]
    =
    \left[\begin{array}{ccc}
        S_{aaa}^{(2)} & S_{aab}^{(2)} & S_{abb}^{(2)} \\
        S_{baa}^{(2)} & S_{bab}^{(2)} & S_{bbb}^{(2)}
    \end{array}\right]
    \left[\begin{array}{c}
        (-\nabla_a T)^2 \\
        2(-\nabla_a T)(-\nabla_b T) \\
        (-\nabla_b T)^2
    \end{array}\right].
\end{equation}
Here, based on the geometry of the generated electric field and the applied temperature gradient, the nonlinear thermoelectric responses can be divided into three, namely, $S_{aaa/bbb}^{(2)}$ the nonlinear Seebeck (NLS) effect, $S_{baa/abb}^{(2)}$ the nonlinear Nernst (NLN) effect and $S_{aab/bab}^{(2)}$ the nonlinear mixed-directional thermoelectric (NLMT) effect. \\

Although bulk WTe$_2$ and TaIrTe$_4$ crystallizes in space group no. 31 ($Pmn$2$_1$), prior electronic \cite{WTe_Torque,WTe_Torque_thickness,TIT_Torque} and magnonic spin \cite{ACS_Nano} transport studies have shown that, in thin flakes, the effective symmetry is reduced such that only the mirror symmetry remains. This reduction has been experimentally confirmed in WTe$_2$/Py heterostructures with thicknesses up to 16 nm through second-harmonic Hall measurements \cite{WTe_Torque_thickness}. In addition, nonlinear Hall (NLH) response has been reported for 29 nm-thick TaIrTe$_4$ and 30 nm-thick WTe$_2$ flakes \cite{kumar2021room}. Since the flakes used in this work have comparable thicknesses, we expect a similar symmetry reduction, with only the mirror symmetry preserved. Exfoliated thin WTe\textsubscript{2} and TaIrTe\textsubscript{4} thus belong to space group no. 6 ($Pm$), and possess only a single mirror symmetry with respect to the $bc$-plane [denoted by \textit{$\sigma(bc)$}]. Under this symmetry operation, the electric field and the applied temperature gradient transform as follows: 

\begin{equation}
\label{Symm_constratints}
    E_i, \nabla_i T \longrightarrow
    \begin{cases}
        -E_i, -\nabla_i T \quad \text{for } i \equiv \hat{a} \text{ under } \sigma(bc), \\
        +E_i, +\nabla_i T \quad \text{for } i \equiv \hat{b}, \hat{c} \text{ under } \sigma(bc).
    \end{cases}
\end{equation}
Imposing the constraints form Eq.~\ref{Symm_constratints} in Eq. \ref{Matrix_form}, allows for the identification of the terms that are forbidden by the crystalline symmetry. Thus, under the constraints of the crystal symmetry of WTe\textsubscript{2} and TaIrTe\textsubscript{4}, Eq. \ref{Matrix_form} reduces to: 

\begin{equation}
\label{Reduced_Matrix_form}
    \left[\begin{array}{c}E_a \\ E_b\end{array}\right]
    =
    \left[\begin{array}{ccc}
        0 & S_{aab}^{(2)} & 0 \\
        S_{baa}^{(2)} & 0 & S_{bbb}^{(2)}
    \end{array}\right]
    \left[\begin{array}{c}
        (-\nabla_a T)^2 \\
        2(-\nabla_a T)(-\nabla_b T) \\
        (-\nabla_b T)^2
    \end{array}\right].
\end{equation}\\

It can be seen that of the 6 independent nonlinear thermoelectric response coefficients, 3 are allowed by the mirror symmetry of the crystal while three are forbidden, namely, $S_{bbb/aab/baa}^{(2)}$ ($\neq$ 0) are allowed by symmetry while $S_{aaa/bab/abb}^{(2)}$ (= 0) are forbidden. \\

Furthermore, we note that when the entire set of crystal symmetry operations of bulk WTe$_2$ and TaIrTe$_4$ are considered including the glide mirror symmetry $[(x,y,z)\rightarrow(x+1/2,-y,z+1/2)]$ and the two-fold screw rotation $[(x,y,z)\rightarrow(-x+1/2,-y,z+1/2)]$, the symmetry constraints permit only the NLMT response, while the NLS and NLN effects are forbidden.  However, for exfoliated thin flakes, under the reduced crystalline symmetry constraints wherein only one mirror symmetry of the crystal persists, all three nonlinear thermoelectric responses, namely the NLMT, NLS and NLN effects are not forbidden by symmetry and therefore can be finite.\\

Although the phenomenological description presented above provides information on which of the nonlinear thermoelectric effects are forbidden by symmetry, they provide no additional information on the microscopic mechanism responsible for these. If the nonlinear thermoelectric effects originate from the Berry curvature dipole (BCD), additional symmetry constraints arise. In a two-dimensional system, the Berry curvature behaves as a pseudoscalar, with only the out-of-plane component being nonzero. Consequently, the BCD acts as an in-plane pseudovector. In two dimensions, the nonlinear thermoelectric response induced by the BCD can generally be expressed as

\begin{equation}
\label{NLN_BCD}
\boldsymbol{E}=\alpha\left[\hat{c} \times \left(-\boldsymbol{\nabla} T\right)\right]\left[\boldsymbol{D} \cdot \left(-\boldsymbol{\nabla} T\right)\right],
\end{equation}

where $\boldsymbol{D}$ denotes the BCD and $\alpha$ is a proportionality coefficient. This expression naturally follows as the thermoelectric counterpart of the NLH effect as described in Ref.~[\onlinecite{sodemann2015quantumSI}].
Rewriting this equation in matrix form, as in Eq.~\ref{Matrix_form}, yields
\begin{equation}
\label{BCD_matrix}
    \left[\begin{array}{c}E_a \\ E_b\end{array}\right]
    = \alpha
    \left[\begin{array}{ccc}
        0 &  D_a & D_b\ \\
        -D_a & -D_b & 0\
    \end{array}\right]
    \left[\begin{array}{c}
        (-\nabla_a T)^2 \\
        2(-\nabla_a T)(-\nabla_b T) \\
        (-\nabla_b T)^2
    \end{array}\right].
\end{equation}

In WTe\textsubscript{2} and TaIrTe\textsubscript{4}, the mirror symmetry $\sigma(bc)$ enforces $D_b$ to be zero. As a result, Eq.~\ref{BCD_matrix} reduces to

\begin{equation}
\label{BCD_matrix_reduced}
    \left[\begin{array}{c}E_a \\ E_b\end{array}\right]
    = \alpha
    \left[\begin{array}{ccc}
        0 &  D_a & 0\ \\
        -D_a & 0 & 0\
    \end{array}\right]
    \left[\begin{array}{c}
        (-\nabla_a T)^2 \\
        2(-\nabla_a T)(-\nabla_b T) \\
        (-\nabla_b T)^2
    \end{array}\right].
\end{equation}

The intrinsic contribution to the nonlinear thermoelectric responses, governed by the BCD, can thus only give rise to the components $S_{baa}^{(2)}$ and $S_{aab}^{(2)}$, subject to the constraint $S_{baa}^{(2)} = -S_{aab}^{(2)}$. The component $S_{bbb}^{(2)}$, namely the NLS effect cannot arise from BCD and thus can arise only from scattering mechanisms.

\subsection{Effect of off-axis thermal biasing on linear thermoelectric responses - Transverse Seebeck Effect}

Although the Seebeck coefficient is often approximated as a scalar that relates the generated electric field collinear to an applied temperature gradient, this simplification is only valid for isotropic materials. In general, the Seebeck coefficient is a second rank tensor that captures the coupling between temperature gradients and the resulting electric field. For anisotropic materials such as WTe\textsubscript{2} and TaIrTe\textsubscript{4}, the Seebeck coefficient depends on crystallographic orientation and differs along the different high symmetry crystallographic axes. For simplicity, omitting one direction namely, $\hat{c}$, the resulting Seebeck tensor for anisotropic materials like WTe\textsubscript{2} and TaIrTe\textsubscript{4} is given by:  

\begin{equation} \label{Seebeck_1}
\binom{E_a}{E_b}
=
\begin{pmatrix}
S_{aa}^{(1)} & 0 \\
0 & S_{bb}^{(1)}
\end{pmatrix}
\binom{-\nabla_a T}{-\nabla_b T},
\end{equation}
where $S_{aa}^{(1)}$ and $S_{bb}^{(1)}$ denote the Seebeck coefficients along the different principal crystallographic axes. The diagonal elements of the Seebeck tensor describe the longitudinal Seebeck response, whereas the off-diagonal elements correspond to the transverse voltage generation. For temperature gradients oriented along a high symmetry crystallographic axis, the generation of a transverse voltage is forbidden by the mirror symmetry of WTe\textsubscript{2} and TaIrTe\textsubscript{4}.\\

For temperature gradients not oriented along one of the high symmetry axes, the response can be estimated by performing a change of the basis in Eq.~\ref{Seebeck_1}  (rotating the basis from $\hat{a}$, $\hat{b}$ to the set of vectors $x$, $y$, rotated by an angle $\gamma$). The resulting Seebeck response tensor for an off-axis temperature gradient is given by: 

\begin{equation} \label{Seebeck_11}
\left[\begin{array}{c}
{E_x}\\{E_y}
\end{array}\right]
=
\left[\begin{array}{cc}
S_{aa}^{(1)} \cos^2 \gamma + S_{bb}^{(1)} \sin^2 \gamma & \cos\gamma \sin\gamma (S_{aa}^{(1)} - S_{bb}^{(1)}) \\
\cos\gamma \sin\gamma (S_{aa}^{(1)} - S_{bb}^{(1)}) & S_{bb}^{(1)} \cos^2 \gamma + S_{aa}^{(1)} \sin^2 \gamma
\end{array}\right]
\left[\begin{array}{c}
{-\nabla_x T}\\{-\nabla_y T}
\end{array}\right].
\end{equation}

Thus, upon rotation of the basis, it can be seen that the transverse components of the Seebeck tensor now become non-zero. For a temperature gradient applied along an off-axis direction $x$, at an angle $\gamma$ with respect to the crystallographic $a$-axis, an additional transverse voltage is generated, namely, $V_y = -\cos \gamma \sin \gamma (S_{aa}^{(1)} - S_{bb}^{(1)})\nabla_x T$, referred to as the transverse Seebeck effect, whose magnitude scales with the difference of the Seebeck coefficients between the two principal axes \cite{goldsmid2011application,tang2015p}. In the case of transverse Seebeck, we note that both the temperature gradient and the voltage response measured normal to the temperature gradient are both oriented away from the high-symmetry crystallographic axis. We emphasize that while the linear Nernst effect arises only under an applied magnetic field, a transverse thermopower proportional to the temperature gradient can nevertheless emerge from the transverse Seebeck response in the second-harmonic (2$\omega$) voltage response induced by a slight misalignment of the WTe\textsubscript{2} or TaIrTe\textsubscript{4} high-symmetry crystal axes with respect to the applied temperature gradient. As we report here, the Seebeck coefficients of WTe\textsubscript{2} along the crystallographic \textit{a}- and \textit{b}-axes exhibit pronounced anisotropy (see Section \ref{Sec:Anisotropic_Response}). Moreover, in bulk TaIrTe\textsubscript{4}, the Seebeck coefficients along different crystallographic directions have been reported to possess opposite signs, reflecting its goniopolar character~\cite{Gonio_TIT}. We thus expect that a misalignment between the applied temperature gradient from high-symmetry axes in WTe\textsubscript{2} or TaIrTe\textsubscript{4} to produce a measurable linear transverse thermopower via the transverse Seebeck effect. This is taken into account throughout our work as mentioned in the main text.
\subsection{Effect of off-axis thermal biasing on nonlinear thermoelectric responses} \label{Sec:Off_Axis_NL}

The nonlinear thermoelectric response in Eq. \ref{Reduced_Matrix_form} can be written in the compact form as: 

\begin{equation*}
\mathbf{E} = \mathbf{M}\,\mathbf{v},
\end{equation*}
where the generated electric fields and the temperature gradients are captured by $\mathbf{E}$ (a 2 $\times$ 1 vector) and $\mathbf{v}$ (a 3 $\times$ 1 vector) respectively, and are related through the nonlinear response matrix $\mathbf{M}$ (a 2 $\times$ 3 matrix). The nonlinear thermoelectric response of the system under an off-axis thermal biasing can be evaluated by rotating the basis from the crystallographic axis to the thermal biasing axes in a similar fashion to that presented for the linear thermoelectric effects in the previous section. For thermal biasing at an angle $\gamma$ with respect to the \textit{a}-axis, the rotation matrix $\mathbf{R}(\gamma)$ denoting the rotation matrix is given by 

\begin{equation*}
\mathbf{R}(\gamma)=
\begin{pmatrix}
\cos\gamma & -\sin\gamma \\
\sin\gamma & \cos\gamma
\end{pmatrix}.
\end{equation*}
The electric field and the temperature gradients transform as: $\mathbf{E}_{ab} = \mathbf{R}\,\mathbf{E}_{xy}, \mathbf{\nabla}_{ab} \mathbf{T} = \mathbf{R}\,\mathbf{\nabla}_{xy} \mathbf{T}$. Denoting $c$ = $\cos \gamma$, $s$ = $\sin \gamma$, the transformation of the temperature gradient is then given by: 

\begin{equation*}
\begin{pmatrix}
-\nabla_a T \\
-\nabla_b T
\end{pmatrix}
=
\begin{pmatrix}
c & -s \\
s & c
\end{pmatrix}
\begin{pmatrix}
-\nabla_x T \\
-\nabla_y T
\end{pmatrix}.
\end{equation*}
For the nonlinear thermoelectric response, defining the quadratic temperature gradients as: 

\begin{equation}
\mathbf{v}_{ab}=
\left[\begin{array}{c}
(-\nabla_a T)^2 \\
2(-\nabla_a T)(-\nabla_b T) \\
(-\nabla_b T)^2
\end{array}\right],
\qquad
\mathbf{v}_{xy}=
\left[\begin{array}{c}
(-\nabla_x T)^2 \\
2(-\nabla_x T)(-\nabla_y T) \\
(-\nabla_y T)^2
\end{array}\right].
\end{equation}
It can be seen that the two vectors are related under the rotation of the basis through $\mathbf{v}_{ab}=\mathbf{Q}(\gamma)\,\mathbf{v}_{xy}$, where

\begin{equation*}
\mathbf{Q}(\gamma)=
\begin{pmatrix}
c^2 & -cs & s^2 \\
2cs & c^2-s^2 & -2cs \\
s^2 & cs & c^2
\end{pmatrix}.
\end{equation*}

The transformation of the nonlinear responses is then given by: 

\begin{equation*}
\begin{split}
\mathbf{E}_{ab} &= \mathbf{M}_{ab}\,\mathbf{v}_{ab}, \\
\mathbf{E}_{xy}
&= \mathbf{R}^T\mathbf{E}_{ab}
= \mathbf{R}^T\mathbf{M}_{ab}\mathbf{v}_{ab}
= \mathbf{R}^T\mathbf{M}_{ab}\mathbf{Q}(\gamma)\mathbf{v}_{xy}.
\end{split}
\end{equation*}

Therefore, the nonlinear response matrix in the rotated basis is given by: 
\begin{equation}
\boxed{
\mathbf{M}_{xy}=\mathbf{R}^T\,\mathbf{M}_{ab}\,\mathbf{Q}(\gamma)
}
\end{equation}

or explicitly,
\begin{equation}
\mathbf{M}_{xy}=
\begin{pmatrix}
m'_{11} & m'_{12} & m'_{13} \\
m'_{21} & m'_{22} & m'_{23}
\end{pmatrix},
\end{equation}

where 

\begin{equation}
    \begin{aligned}
        m_{11}^{'}  = s \left[ 2S_{aab}^{(2)} c^2 + S_{baa}^{(2)} c^2 + S_{bbb}^{(2)} s^2  \right] \quad &\text{,} \quad 
         m_{21}^{'}  = c \left[ -2S_{aab}^{(2)} s^2 + S_{baa}^{(2)} c^2 + S_{bbb}^{(2)} s^2  \right], \\ 
        m_{12}^{'} = c \left[ S_{aab}^{(2)} (c^2 - s^2) - S_{baa}^{(2)} s^2 + S_{bbb}^{(2)} s^2  \right] \quad &\text{,} \quad 
         m_{22}^{'}  = s \left[ -S_{aab}^{(2)} (c^2 - s^2) - S_{baa}^{(2)} c^2 + S_{bbb}^{(2)} c^2  \right], \\
        m_{13}^{'} = s \left[ -2S_{aab}^{(2)} c^2 + S_{baa}^{(2)} s^2 + S_{bbb}^{(2)} c^2  \right] \quad &\text{,} \quad 
        m_{23}^{'}  = c \left[ 2S_{aab}^{(2)} s^2 + S_{baa}^{(2)} s^2 + S_{bbb}^{(2)} c^2  \right]. \\
    \end{aligned}
\end{equation}

Similar to the linear thermoelectric response, wherein a transverse voltage generation is symmetry-forbidden when the temperature gradient is oriented along a high-symmetry crystallographic axis, but becomes finite when the thermal bias is oriented at an off-axis direction (due to the transverse Seebeck effect) in the absence of a magnetic field, the nonlinear thermoelectric responses exhibit an analogous behavior. Specifically, although the coefficients corresponding to m$_{11/13/22}$ (which corresponds to $S_{aaa/abb/bab}^{(2)}$) vanish when the temperature gradient is oriented along a high-symmetry axis, they acquire finite values when the temperature gradient is oriented away from the high symmetry crystallographic axes.

\section{Low-Frequency transport measurements with Lock-In Detection}

The voltage response of the system $\left[V_{in}(t)\right]$ for an ac driving current $[I(t)$ = $\sqrt{2}I_{0}$ $\sin(\omega t)$, where $I\textsubscript{0}$ is the rms amplitude of the current$]$ can be expressed as:

\begin{equation*}
    V_{in}(t)=\sum_{j=1}^{\infty} R_j I^j(t),
\end{equation*}
where $R_n$ denotes the $n^{\mathrm{th}}$-order response of the system to the current $I(t)$. The distinct n\textsuperscript{th}-order responses generated in response to the driving ac current are measured using a lock-in amplifier, which multiplies the measured voltage by a reference signal and performs time averaging to selectively extract the response at the $n^{\mathrm{th}}$ multiple of the driving frequency ($n$\textsuperscript{th} harmonic). The $n^{\mathrm{th}}$ harmonic voltage response $[V_n(t)]$ can be expressed as:

\begin{equation*}
V_n(t)=\frac{\sqrt{2}}{T} \int_{t}^{t+T} \sin (n \omega s+\phi) V_{\text {in }}(s) d s,
\end{equation*}
where $\phi$ = 0$^{\circ}$ for odd-$n$ and $\phi$ = -90$^{\circ}$ for even-$n$. By substitution of the input voltage, the $n^{\mathrm{th}}$ harmonic lock-in voltage responses can be expressed as: 

\begin{equation}
\label{Eq:Lockin}
    \begin{aligned}
        V_1 & = I_0R_1 + \frac{3}{2}I^3_0R_3 \quad \text{,} \quad 
        V_2  = \frac{1}{\sqrt{2}}\left[ I^2_0R_2 + 2I^4_0R_4 \right], \\ 
        V_3 & = -\frac{1}{2}I^3_0R_3 \quad \text{,} \quad 
        V_4  = -\frac{1}{2\sqrt{2}}\left[I^4_0R_4 + 3 I^6_0R_6 \right],\\
        V_6  &= \frac{1}{4\sqrt{2}}I^6_0R_6.
    \end{aligned}
\end{equation}

To measure the linear and the nonlinear thermoelectric effects, a temperature gradient is generated by driving an ac current through the heater electrode, [$I$ = $\sqrt{2}I_{0}$ $\sin(\omega t)$]. The generated temperature gradient (from Joule heating of the heater) due to the ac current can be expressed as: 

\begin{equation*}
    \begin{aligned}
    \nabla T & = A I^2(t) R, \\
    & = (A R_\mathrm{Heater}) (\sqrt{2}I_0\sin(\omega t))^2, \\ 
    \end{aligned}
\end{equation*}
where R$_\mathrm{Heater}$ is the resistance of the heater electrode. Thus, the nonlinear thermoelectric responses, which depend quadratically on the temperature gradient, are detected as the fourth-harmonic voltage responses with respect to the driving frequency of the ac current in the heater. To extract the nonlinear thermoelectric coefficients from the measured voltage responses, we thus take into account the constants in Eq. \ref{Eq:Lockin}, for instance, for the case of nonlinear Nernst effect: 

$$
\begin{aligned}
V_b^\text{NNE} & =L S_{baa}^{(2)}\left(C_0^2 I^4(t)\right) \Rightarrow R_4 \equiv C_0^2 L S_{baa}^{(2)}. \\
\end{aligned}
$$

We perform $V$–$I$ measurements as a function of the heater current $I_{0}$ and the reported voltage responses for $V_n$ versus $I_0$ corresponds to the $n^{\mathrm{th}}$-harmonic voltage $V_{n}$ measured while increasing the rms amplitude of the heater current. For the nonlinear thermoelectric effects for instance, the slope of $V^{4\omega}$ plotted against $I_{0}^{4}$ yields $-\frac{R_{4}}{2\sqrt{2}}$. We note that the associated numerical prefactors are accounted for in extracting both the linear and nonlinear thermoelectric coefficients from the measured lock-in voltage responses.

\section{Characterization of the devices}

The devices measured in this work include WTe$_{2}$ and TaIrTe$_4$ devices used to probe nonlinear thermoelectric responses in the NLS, NLN and NLMT configurations as presented in the main text. To verify the role of crystalline symmetry, we additionally fabricated two WTe$_{2}$ devices to investigate the symmetry-forbidden nonlinear thermoelectric responses in the NLN and NLMT configurations. Furthermore, two additional WTe$_{2}$ devices were fabricated - one to investigate the anisotropy in the linear thermoelectric effects (referred to as WTe$_2$ (Anisotropy)) and one in a Hall bar geometry to measure the nonlinear Hall effect (referred to as WTe$_2$ (Hall Bar)). To investigate the role of the crystalline symmetry, it is necessary to vary both the crystallographic direction for measuring the voltage response and the direction of thermal biasing that requires using a separate device for each specified configuration. The crystal axes orientation of the WTe$_{2}$ and TaIrTe$_4$ flakes measured were determined from polarized Raman measurements based on previously reported analogous measurements \cite{ACS_Nano, kim2016determination, liu2018raman}.

\subsection{Polarized Raman measurements of the devices} \label{Sec:Raman}

The polarized Raman spectra were obtained with an inVia Raman Renishaw microscope utilizing a linearly polarized laser in back-scattering geometry. The excitation wavelength and the grating used were $\lambda$ = 532 nm and 1800 l/mm, respectively. The laser power was kept below $\SI{1}{mW}$ with a diffraction-limited spot of $\sim \SI{1}{\micro m}$ positioned on the WTe$_{2}$ and TaIrTe$_4$ flakes of the devices. The samples were rotated in steps of $\sim$10$^{\circ}$ or $\sim$15$^{\circ}$ (depending on the sample) for each measurement. Optical images were recorded before each measurement to determine the exact rotation between the measurements.\\

\begin{figure}[htb]
   \centering
\includegraphics[width=\textwidth]{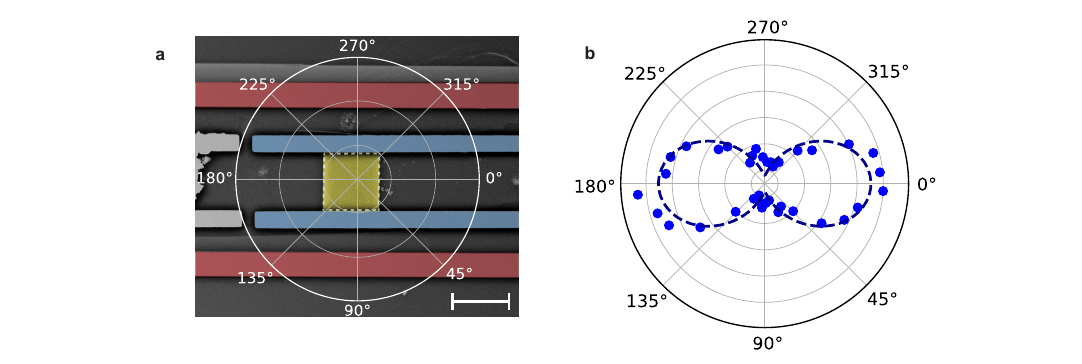}  
   \caption{Representative measurement of Polarized Raman spectroscopy used to determine the crystal axes. \textbf{a}, A false-coloured SEM image of the TaIrTe$_4$ device fabricated to measure the NLS effect. TaIrTe$_4$, Pt heaters, and Au contacts are shown in yellow, red, and light blue, respectively (scale bar: $\SI{5} {\micro\meter}$). \textbf{b}, The intensity of the Raman peak corresponding to $\sim$147.7 cm$^{-1}$ plotted as a function of the sample rotation angle. The angle is as defined in panel \textbf{a}. The dashed line represents a fit to a $\cos^2 \gamma$ function.}
   \label{fig_Raman}
\end{figure}

For WTe$_2$, the Raman peaks at $\sim$165.7 cm$^{-1}$ and $\sim$211.3 cm$^{-1}$ correspond to singly degenerate vibrational modes~\cite{kim2016determination}. The peak at $\sim$165.7 cm$^{-1}$ exhibits maximum intensity along the crystallographic $a$-axis, whereas the peak at $\sim$211.3 cm$^{-1}$ shows maximum intensity along the $b$-axis. Therefore, the ratio of the intensities of the $\sim$165.7 cm$^{-1}$ and $\sim$211.3 cm$^{-1}$ peaks were used to determine the device orientation. By fitting this ratio with a $\cos^2(\gamma)$ function, the $a$-axis was identified.
For TaIrTe$_4$, the Raman peak at $\sim$147.7 cm$^{-1}$ corresponds to a singly degenerate vibrational mode and exhibits maximum intensity along the crystallographic $a$-axis~\cite{liu2018raman}. By fitting the angular dependence of this peak intensity to a $\cos^2(\gamma)$ function, we determine the crystallographic a-axis of our devices, as shown in Fig.~\ref{fig_Raman} for the NLS device. Additionally, the geometrical dimensions and the thickness' of the flakes in the devices measured were determined by scanning electron microscopy and atomic force microscopy respectively. The offsets from the assumed crystallographic orientation and the thickness' of each of the flakes used in the various devices are summarized in Table \ref{tab:Summary}.\\

\begin{table}[H]
\centering
\begin{tabular}{|c|c|c|}
\hline
\textbf{Device Geometry} & \textbf{Offset Angle ($\gamma$)$^{*}$} & \textbf{Thickness} \\
\hline
WTe$_2$ (NLN) & $\gamma_{b}$ $<$ 0.5$^\circ$ & $\sim$ 32.6 nm \\ \hline
WTe$_2$ (NLS)& $\gamma_{b}$ $\sim$ 2.5$^\circ$ & $\sim$ 27 nm \\ \hline
WTe$_2$ (NLMT)& $\gamma_{a}$ $\sim$ 1$^\circ$ & $\sim$ 20.4 nm  \\ \hline
WTe$_2$ (Forbidden - NLN)& $\gamma_{b}$ $<$ 2.5$^\circ$  & $\sim$ 15 nm \\ \hline
WTe$_2$ (Forbidden - NLMT)& $\gamma_{a}$ $<$ 2.5$^\circ$  & $\sim$ 18.1 nm \\ \hline
WTe$_2$ (Anisotropy)& $\gamma$ $\sim$ 3$^\circ$ & \makecell {$\sim$ 14 nm (: samples for $\nu_{ba}$ and $S_{aa}^{(1)}$ measurements)\\
$\sim$ 17 nm (: samples for $\nu_{ab}$ and $S_{bb}^{(1)}$ measurements)} \\ \hline
WTe$_2$ (Hall Bar)& $\gamma$ $<$ 0.5$^\circ$ & $\sim$ 15.6 nm \\ \hline
TaIrTe$_4$ (NLN)& $\gamma_{b}$ $\sim$ 1$^\circ$ & $\sim$ 30 nm \\ \hline
TaIrTe$_4$ (NLS)& $\gamma_{b}$ $<$ 0.5$^\circ$ & $\sim$ 19.5 nm \\ \hline
TaIrTe$_4$ (NLMT)& $\gamma_{a}$ $\sim$ 1$^\circ$ & $\sim$ 7.4 nm \\ \hline
\end{tabular}

\begin{tablenotes}[para]
\footnotesize
\centering
\item[*] $\gamma_{a/b}$ denotes the offset of the voltage detection channel from the crystallographic-\textit{a} and -\textit{b} axes respectively while $\gamma$ denotes the offset of the voltage detection channel with respect to the high symmetry crystallographic axes in devices that employ voltage detection along both crystallographic axes.
\end{tablenotes}

\caption{Summary of the offsets between the high symmetry crystallographic axes and the channel of voltage detection and the thickness' of the flakes measured in the respective devices.}
\label{tab:Summary}
\end{table}


\section{Linear thermoelectric coefficients of the NLS and NLN devices} \label{Linear_Thermo}

\begin{figure}[htb]
   \centering
\includegraphics[width=\textwidth]{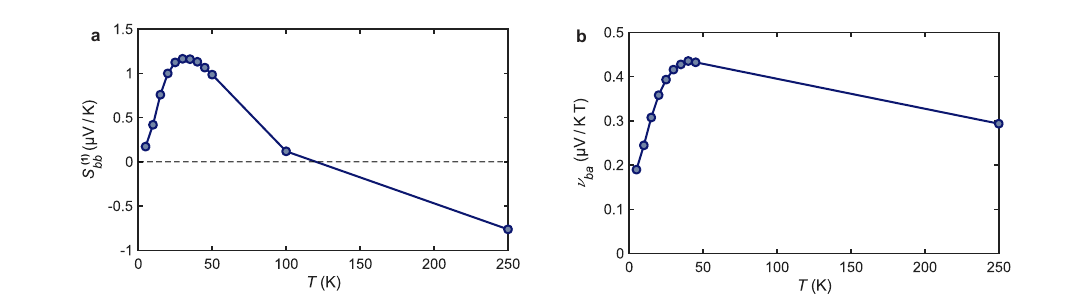}   
   \caption{Linear thermoelectric coefficients of WTe\textsubscript{2}. \textbf{a}, Temperature dependence of the Seebeck coefficient for $\nabla_bT$ oriented along the crystallographic $b$-axis. \textbf{b}, Temperature dependence of the Nernst coefficient ($\nu_{ba}$), defined by $E_b = \nu_{ba} \nabla_a T\ B_c$, where $B_c$ denotes the magnetic field in the out-of-plane direction. Error bars are smaller than the data points.}
   \label{fig_Lin_WT}
\end{figure}

 To compare the linear and the nonlinear thermoelectric effects, we extract the linear thermoelectric coefficients, namely the Seebeck and Nernst coefficients, as shown in Figs.~\ref{fig_Lin_WT}\textbf{a} and \textbf{b} from the devices reported in the main text for WTe$_2$. The Seebeck coefficient increases from zero to positive values at low temperatures, reaching a maximum around 40 K. At higher temperatures, the response diminishes and becomes negative at temperatures above approximately 100 K, consistent with previous reports \cite{rana2018thermopower, wu2015temperature}. For the Nernst coefficient (measured in the presence of a magnetic field), a positive maximum is observed around 40 K, after which it decreases with increasing temperature but remains positive, qualitatively following the trend reported previously~\cite{pan2022ultrahigh}. The pronounced linear thermoelectric responses, particularly at low temperatures, reflect the semi-metallic character of the material and its high carrier mobilities for both electrons and holes~\cite{pan2022ultrahigh}. Furthermore we observe that the linear thermoelectric effects are anisotropic and depend on the crystallographic axis and analyze this anisotropy in detail in Section \ref{Sec:Anisotropic_Response}. 
 \\

\begin{figure}[H]
   \centering
   \includegraphics[width=\textwidth]{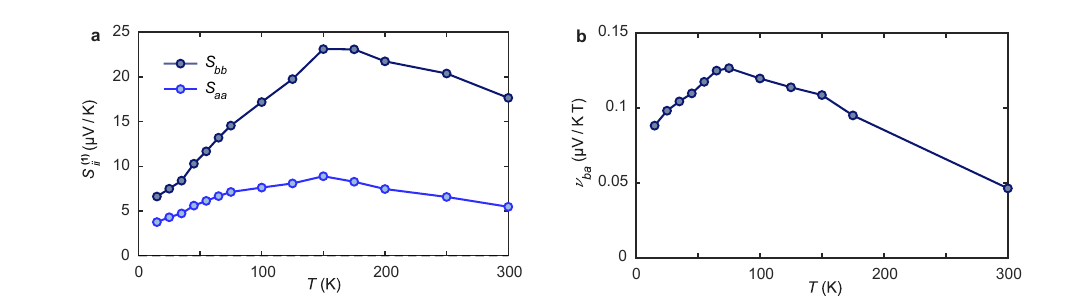} 
   \caption{Linear thermoelectric coefficients of TaIrTe\textsubscript{4}. \textbf{a}, Temperature dependence of the Seebeck coefficient along the crystallographic \textit{a}- and \textit{b}-axes. \textbf{b}, Temperature dependence of the Nernst coefficient  ($\nu_{ba}$), corresponding to $E_b = \nu_{ba} \nabla_a T\ B_c$, where $B_c$ denotes the magnetic field applied oriented out-of-plane.
   Error bars are smaller than the data points.}
   \label{fig_Lin_TIT}
\end{figure}

The linear thermoelectric coefficients of TaIrTe$_4$, extracted from the devices described in the main text, are presented in Fig. \ref{fig_Lin_TIT}. In contrast to Ref. \citen{Gonio_TIT}, wherein the goniopolar character of TaIrTe$_4$ gives rise to Seebeck coefficients of opposite signs along the principal crystallographic axes, namely that the $S^{(1)}_{aa}$ and $S^{(1)}_{bb}$ are of opposite signs while we instead observe the same sign along both crystallographic directions. A plausible explanation is the unintentional doping of the unencapsulated flakes during device fabrication. Within the Mott framework, the Seebeck coefficient is related to the conductivity through: $\mathbf{S^{(1)}} = -\frac{\pi^2}{3}\frac{k_B^2 T}{e}\left.\frac{\partial \ln \boldsymbol{\sigma}}{\partial \varepsilon}\right|_{\varepsilon_\mathrm{F}}$. Notably, the Seebeck coefficient is governed by the energy derivative of the conductivity at the Fermi level; a doping-induced shift of $\varepsilon_\mathrm{F}$ can therefore modify both the magnitude and sign of $\mathbf{S^{(1)}}$. To independently verify the crystallographic orientation of the voltage detection channel, we perform polarization-resolved Raman spectroscopy (see SM \ref{Sec:Raman}).

\section{Anisotropic Electrical and Thermoelectric Transport} \label{Sec:Anisotropic_Response}

\subsection{Resistivity anisotropy in WTe$_2$}

\begin{figure}[H]
    \centering
        \makebox[\textwidth][c]{%
    \includegraphics[width=1.07\textwidth]{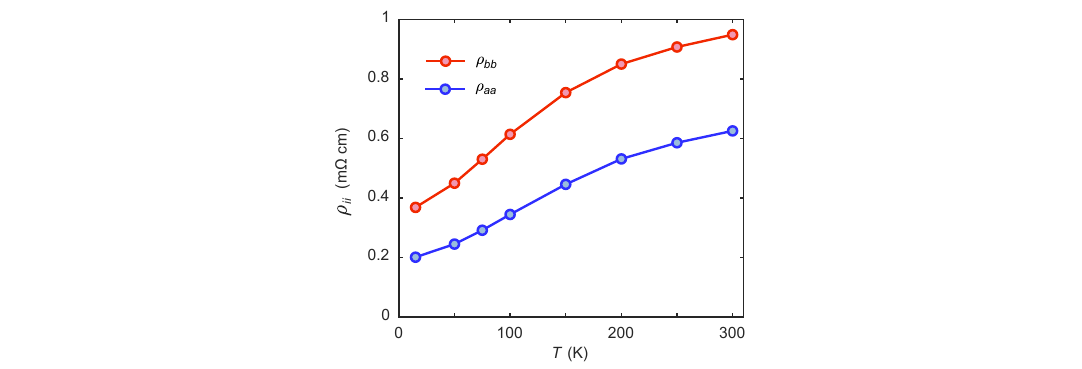}   
    }
    \caption{The four-probe longitudinal electrical resistivity of WTe$_2$ measured using the WTe$_2$ Hall bar devices along the crystallographic \textit{a-} and \textit{b-} axes as a function of temperature. Error bars are smaller than the data points.}
    \label{fig_res_anisotropy}
\end{figure}

\subsection{Anisotropic linear thermoelectric responses in WTe$_2$}

\begin{figure}[h]
    \centering
        \makebox[\textwidth][c]{%
    \includegraphics[width=1.07\textwidth]{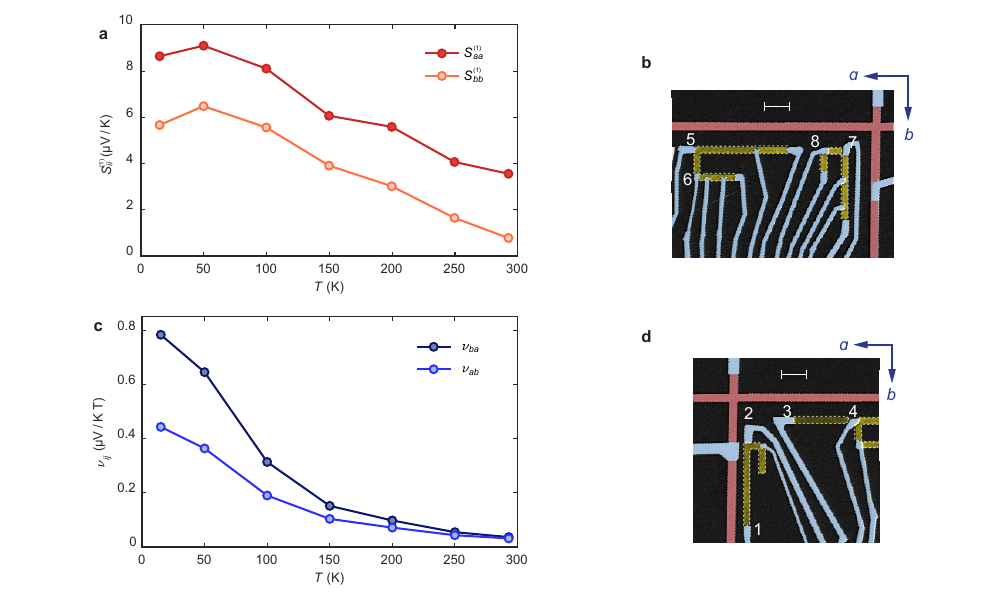}   
    }
    \caption{\textbf{a}, The Seebeck coefficient of WTe$_2$ measured for the two different crystallographic directions; $S_{aa}^{(1)}$ and $S_{bb}^{(1)}$. \textbf{b}, False-coloured SEM image of the device used for the measurement of the Seebeck coefficient shown in panel \textbf{a} (Scale bar: $\SI{10} {\micro\meter}$). $S_{aa}^{(1)}$ was measured using the voltage probes 7-8 and the upper heater electrode while $S_{bb}^{(1)}$ was measured using the voltage probes 5-6 and the heater electrode on the right side. \textbf{c}, The Nernst coefficient measured for when the temperature gradient is oriented along the the two different crystallographic directions; $\nu_{ba}$ and $\nu_{ab}$. \textbf{d}, False-coloured SEM image of the device used for the measurement of the Nernst coefficient shown in panel \textbf{c} (Scale bar: $\SI{10} {\micro\meter}$). $\nu_{ba}$ was measured using the voltage probes 1-2 and the upper heater electrode while $\nu_{ab}$ was measured using the voltage probes 3-4 and the heater electrode on the left side. Error bars are smaller than the data points.}
    \label{fig:anisotropic_linear}
\end{figure}

We verify the anisotropic thermoelectric transport in WTe$_2$ through measurements of the Seebeck and Nernst coefficients, as shown in Fig. \ref{fig:anisotropic_linear}. In contrast to the results presented in Fig.~\ref{fig_Lin_WT} for the Seebeck coefficient, wherein we observe a change in the sign of the Seebeck coefficient above 100 K, the Seebeck coefficient in the present device remains positive throughout the entire temperature range below room temperature. We thus emphasize that the thermoelectric response of exfoliated WTe$_2$ can vary significantly from device to device, with even the dominant charge carrier type differing between samples as a function of temperature. One possible explanation for this behavior is unintended doping introduced during the sample fabrication process in these unencapsulated flakes as mentioned earlier for TaIrTe$_4$.\\

In addition, all devices used for these measurements were fabricated on Si/SiO$_2$ substrates. In this case we note that the induced temperature gradient across the flakes is relatively small, especially for low temperatures (see Sec. \ref{Sec:simu_Si}). As a result, the extraction of the absolute thermoelectric coefficients becomes subject to larger uncertainties in this temperature range. This reduced temperature gradient also leads to a temperature profile that differs from that observed for the devices fabricated on fused silica substrates. The Nernst coefficient measurements were performed for magnetic fields below 500 mT. We define the anisotropy in the Nernst coefficient as the difference in the transverse voltage generated for a given temperature gradient and an applied out-of-plane magnetic field, when the directions of thermal biasing and voltage detection are varied relative to the principal crystallographic axes, namely $\nu_{ab}$ and $\nu_{ba}$. \\

We note that the reported anisotropy in the Nernst coefficients seems to appear to violate the Onsager reciprocity relations \cite{Onsager_A}. In particular, from the relation $\mathbf{E}=-\mathbf{S}(B)\nabla T$, one might expect that $S_{ba}(+B)=S_{ab}(-B)$, namely that there is no anisotropy in the Nernst coefficients. We emphasize that this apparent discrepancy arises because the Onsager relations strictly apply to the electrical ($\boldsymbol{\sigma}$) conductivity tensor, rather than directly to the Seebeck (or specifically the magneto-thermopower) tensor \cite{Application_Onsager,Buttiker_TenFold}. Under open-circuit conditions the Seebeck tensor is given by $\mathbf{S}=-\boldsymbol{\rho}\boldsymbol{\alpha}$, where $\boldsymbol{\rho}=\boldsymbol{\sigma}^{-1}$. Consequently,
\begin{align*}
S_{ab}(B) &= -\left[\rho_{aa}(B)\alpha_{ab}(B)+\rho_{ab}(B)\alpha_{bb}(B)\right], \\
S_{ba}(B) &= -\left[\rho_{bb}(B)\alpha_{ba}(B)+\rho_{ba}(B)\alpha_{aa}(B)\right].
\end{align*}

Since these elements involve different combinations of the resistivity and thermoelectric conductivity tensor elements, owing to the anisotropy in the electrical resistivity, $S_{ab}(B)$ and $S_{ba}(B)$ need not be identical. Therefore, an anisotropy in the Nernst coefficient does not constitute a violation of the Onsager reciprocity relations.\\

While the anisotropy of the Seebeck coefficients in WTe$_2$ along the different crystallographic axes has previously been discussed from a theoretical perspective \cite{WTe2_Seebeck_DFT}, here we report the experimental observation of this anisotropic response in both the Seebeck and Nernst coefficients of WTe$_2$. Anisotropy of the Seebeck coefficient in TaIrTe$_4$ has also been reported previously in Ref. \citen{Gonio_TIT}. As discussed in detail in Sec. \ref{Sec:Symm}, a slight misalignment of the crystal axes combined with the anisotropic Seebeck coefficients, can thus lead to the generation of a transverse thermopower through the transverse Seebeck effect. The transverse Seebeck effect thus produces a measurable voltage response at 2$\omega$, even in the geometry for the measurement of nonlinear Nernst as presented in the main text.  

\section{Separation of the intrinsic nonlinear thermoelectric effects from temperature induced nonlinearities}

\subsection{Temperature induced nonlinearities from linear thermoelectric effects}

In addition to the intrinsic nonlinear thermoelectric effects as described in Section \ref{Sec:Symm}, the temperature dependence of the linear thermoelectric coefficients themselves can generate higher-order responses (with respect to the driving ac heater current), including a measurable response at 4$\omega$. The origin of this term can be understood from a Taylor expansion of the temperature dependence of the linear Seebeck coefficient in the Seebeck voltage generated as given below (corresponding to Eq. 3 of the main text):

\begin{equation}
\label{NLE_Linear}
    \begin{aligned}
    +\frac{V_i}{L} & =-S^{(1)}_{ij}(T) \nabla_j T \\ 
    & =-\left(S^{(1)}_{ij}\left(T_0\right)+\left.\frac{\partial S^{(1)}_{ij}}{\partial T}\right|_{T_0} \cdot \Delta T\right) \nabla_{j} T \quad, \\ 
    \end{aligned}
\end{equation}
where both $\Delta T$ and $\nabla_i T$ both depend quadratically on the driving current and thus generate a measurable response at 4$\omega$. Under a driving ac current, there are two factors to account for, namely the rise in temperature of the flake and the temperature gradient across the flake. It is important to note that the temperature rise, namely $\Delta T$ is a scalar quantity - it reflects only the magnitude of the local rise in temperature due to a driving ac current and is independent of the directionality of $\nabla_i T$. In contrast, the temperature gradient $\nabla_iT$ is a vector that determines the sign of the thermoelectric responses. Furthermore, we confirm the functional form such a contribution in the 4$\omega$ voltage response ascribed to the temperature dependence of the linear thermoelectric coefficients utilizing the device comprised of a Hall bar and heaters and is presented in detail in Sec. \ref{Sec:Func_Form}.\\

For instance, in the longitudinal configuration, where the temperature gradient and the resulting electric field are collinear, both the sign of the Seebeck voltage generated and that of the nonlinear thermoelectric response arising from the temperature dependence of the linear Seebeck coefficient are determined by $\nabla_i T$. We emphasize that while $\Delta T$ characterizes the magnitude of the temperature increase, the sign (the directionality) of the resulting nonlinear thermoelectric signal is governed solely by $\nabla_i T$.
\subsection{Separation of the intrinsic nonlinear thermoelectric responses from the temperature-dependent linear thermoelectric effects}

Eq.~\ref{NLE_Linear} can be further rewritten in terms of the driving current of the heater as follows: 

\begin{equation} \label{Eq:Analysis}
    \begin{aligned}
    +\frac{V_i}{L} & =-\left(S_{ij}^{(1)}\left(T_0\right)+\left.\frac{\partial S_{ij}^{(1)}}{\partial T}\right|_{T_0} \cdot \Delta T\right) \nabla_j T \quad, \\ 
    & =-\left(S_{ij}^{(1)}\left(T_0\right)+\left.\frac{\partial S_{ij}^{(1)}}{\partial T}\right|_{T_0}  A_0 I^2(t) \right)\left(A_1 I^2(t)\right) \quad , \\ 
     & =- S_{ij}^{(1)}\left(T_0\right) \cdot A_1 \space I^2(t) -\left.\frac{\partial S_{ij}^{(1)}}{\partial T}\right|_{T_0}  A_0A_1 \space I^4(t) \quad, \\ 
    \end{aligned}
\end{equation}

\begin{figure}[htb]
    \centering
    \includegraphics[width=\textwidth]{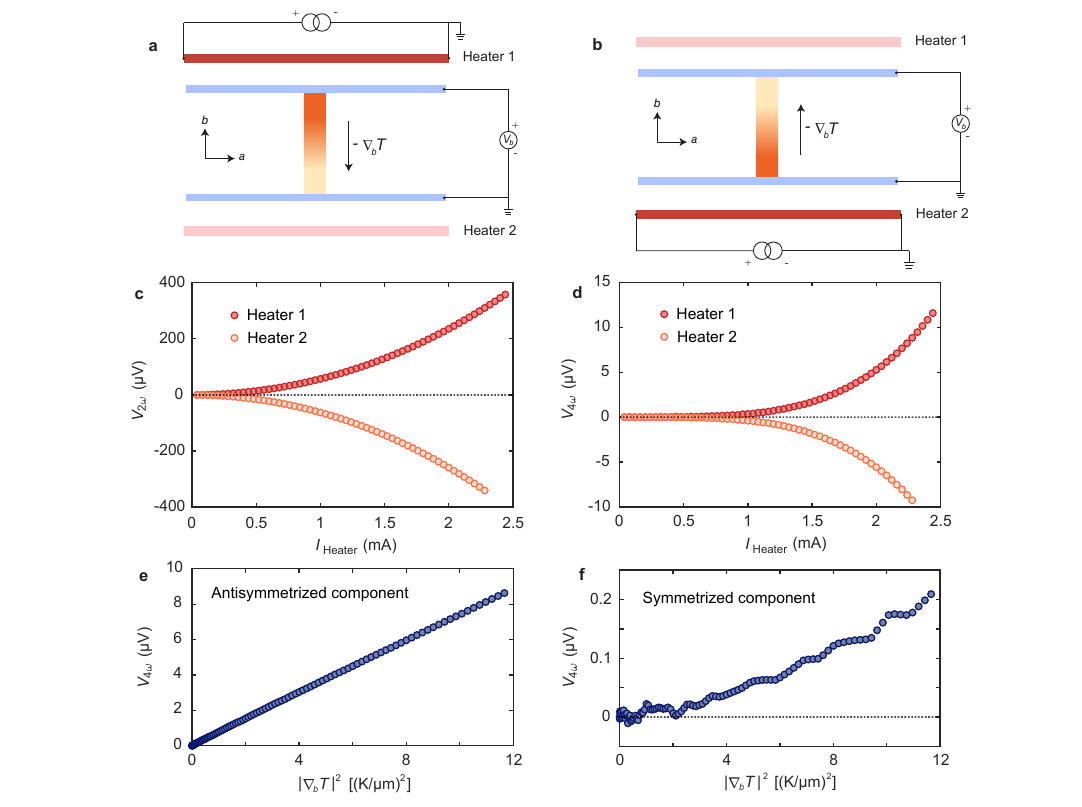} \caption{\textbf{a}-\textbf{b}, Schematic illustrations of the the device geometry for the NLS measurement, employing two heaters on either side of the flake. \textbf{c}-\textbf{d}, The second-harmonic voltage (\textbf{c}) and the fourth-harmonic voltage (\textbf{f}) responses of the TaIrTe$_4$ NLS device at 292.5 K, measured by generating a temperature gradient by driving an ac current in the two different heater electrodes. \textbf{e}-\textbf{f}, Antisymmetrized (\textbf{e}) and the symmetrized (\textbf{f}) fourth-harmonic voltage response from panel \textbf{d}. We note that the (anti-)symmetrization of the fourth-harmonic voltage is performed with respect to the ac currents in the heater electrodes after accounting for the difference in the heater powers extracted by normalizing with the magnitude of the second-harmonic voltage (namely the Seebeck response) shown in panel \textbf{c}.}
    \label{fig:Decouple_Intrinsic}
\end{figure}

where $A_0$ and $A_1$ are proportionality constants that describe the effective temperature rise and the increase of the temperature gradient across the flake in response to the driving ac current in the heater electrode, respectively. As emphasized earlier, only the temperature gradient $\nabla_iT$ determines the sign of the resulting nonlinear thermoelectric voltage response arising from the nonlinearities associated with the temperature dependence of linear thermoelectric coefficients. Consequently, the contribution from the intrinsic nonlinear thermoelectric response can be isolated by using two heaters that effectively reverse the direction of $\nabla_iT$, allowing the temperature-induced nonlinearities to change sign while the intrinsic contribution remains unchanged.
\\

For the dual-heater configuration shown in Fig.~\ref{fig:Decouple_Intrinsic}\textbf{a, b}, the linear and nonlinear thermoelectric responses corresponding to that generated by driving an ac current in the top heater (H1) and the bottom heater (H2) electrodes are given by (taking into account the sign of the Seebeck coefficient as $S$ = -$\left[\frac{V_{Hot} - V_{Cold}}{T_{Hot}-T_{Cold}}\right]$):

\begin{equation}
\label{Eq:H1}
    \begin{aligned}
    \frac{V_{2\omega}^{H1}}{L} & = -S_{bb}^{(1)}|\nabla_b T| \quad,\\
    \frac{V_{4\omega}^{H1}}{L} & = -\left(\left.\frac{\partial S_{bb}^{(1)}}{\partial T}\right|_{T_0} \cdot \Delta T\right) |\nabla_b T| + S_{bbb}^{(2)} (\nabla_b T)^2 \quad,
    \end{aligned}
\end{equation}

\begin{equation}
\label{Eq:H2}
    \begin{aligned}
    \frac{V_{2\omega}^{H2}}{L} & = +S_{bb}^{(1)}|\nabla_b T| \quad,\\
    \frac{V_{4\omega}^{H2}}{L} & = +\left(\left.\frac{\partial S_{bb}^{(1)}}{\partial T}\right|_{T_0} \cdot \Delta T\right) |\nabla_b T| + S_{bbb}^{(2)} (\nabla_b T)^2 \quad.
    \end{aligned}
\end{equation}

It can be seen from Eq.~\ref{Eq:H1} and \ref{Eq:H2} that the contribution from the intrinsic nonlinear thermoelectric responses can be isolated from the temperature-induced nonlinearities by symmetrizing (with respect to the temperature gradients) the 4$\omega$ voltage response for the different heater positions. We note that symmetrizing yields the intrinsic nonlinear thermoelectric response as the convention for voltage detection (namely that the grounding terminal) remains the same between the two configurations. In the case of the reversal of the convention for voltage detection upon changing the heater position, the same can be obtained from the anti-symmetrization of the voltage response with respect to the heater position. We note that unlike in the case of nonlinear electronic transport measurements, where the reversal of the charge current is possible with a given pair of contacts, to reverse the temperature gradient, a geometry with two separate heaters is necessary. We emphasize that this analysis of isolating the intrinsic nonlinear contribution from the temperature dependence of the linear effect is valid for all three configurations shown in Fig.~1\textbf{b}-\textbf{d} in the main text (namely for all elements of $S_{ijk}^{(2)}$), including the geometry used to measure the nonlinear Nernst effect wherein we attribute the second-harmonic voltage response to the transverse Seebeck effect as explained earlier. We now outline the exact procedure used to isolate the intrinsic nonlinear thermoelectric response from the measured V-$I_{0}$ responses from swapping the active heater.\\
\subsection*{Analysis procedure for separating the intrinsic nonlinear thermoelectric response}

To extract the intrinsic nonlinear thermoelectric responses from the measured voltage responses using different active heater electrodes, it is essential to account for the variations in the heaters' current-to-thermal gradient conversion, specifically the differences in the coefficients A$_{i}$ in Eq. \ref{Eq:Analysis}. The simultaneously measured 2$\omega$ voltage response, which corresponds to the linear thermoelectric voltage, allows us to estimate the differences in the thermal gradient generated by the different heater electrodes for a given bias current. By analyzing the ratio of the measured 2$\omega$ voltage response while varying the position of the active heater, we can account for the differences in the current-to-thermal gradient conversions between the two heater electrodes. Upon determining this normalization ratio, we normalize the measured fourth-harmonic voltage response to account for this difference. After this normalization, symmetrizing with respect to the thermal gradient (i.e., the relative position of the active heater) is achieved by symmetrizing with respect to the heater currents which yields the intrinsic nonlinear thermoelectric voltage response. Alternatively phrased, we extract the difference in the measured fourth-harmonic voltage response for a given magnitude of the linear thermoelectric voltage as a function of the heater currents for varying heater positions. We emphasize that this procedure differs however from simultaneously driving both heaters with an ac current to null the linear thermoelectric voltage and probe the fourth-harmonic voltage response. In this case, concurrent heating primarily raises the sample temperature, namely $\Delta$T while reducing the thermal gradient across the flake, namely $\nabla$T, that drives the intrinsic nonlinear thermoelectric responses.\\

We emphasize that by employing both heaters, we are able to isolate the intrinsic nonlinear thermoelectric responses from temperature-induced nonlinearities arising from the linear effects. From Fig.~\ref{fig:Decouple_Intrinsic}\textbf{e, f}, it can be observed that the contribution of the temperature dependent linear thermoelectric effects to the measured fourth-harmonic is about 40 times larger in magnitude in comparison to the intrinsic nonlinear Seebeck effect. \\

In this analysis, we attribute the intrinsic nonlinear thermoelectric voltage response to the in-plane temperature gradients. We note that with a sizable out-of-plane temperature gradient, the symmetry allowed nonlinear Seebeck effect $S_{bbb}^{(2)}$, may also be mimicked by other mixed thermoelectric nonlinear responses corresponding to the tensor components such as $S_{bbc}^{(2)}$ or $S_{bcc}^{(2)}$, both of which are also allowed under the constraints of the crystal symmetry. In addition, the 4$\omega$ voltage response attributed to the nonlinear Seebeck response can also arise from the temperature dependence of the linear Seebeck coefficient $S_{bc}$. We note that these contributions would generate a 4$\omega$ voltage response with the same sign even when the direction of the in-plane temperature gradient is reversed by swapping the relative position of the active heater. We estimate that the temperature gradient along the out-of-plane direction is approximately four orders of magnitude smaller than that within the \textit{ab}-plane, supporting the assumption that the observed nonlinear thermoelectric voltage response originates from in-plane temperature gradients (for details see Sec.~\ref{Sec:Grad_OOP}).\\


\subsection*{Effect of misalignment-induced nonlinear thermoelectric signals that does not reverse with heater switching}

\begin{figure}[htb]
    \centering
    \includegraphics[width=\textwidth]{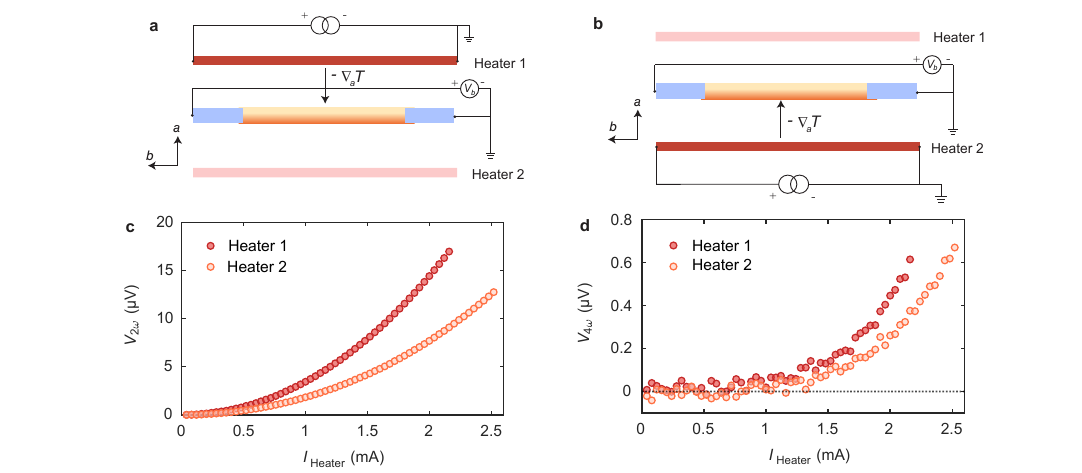} \caption{\textbf{a}-\textbf{b}, Schematic illustration of the device geometry used for the NLN measurement, employing two heaters on either side of the flake. \textbf{c}-\textbf{d}, The second-harmonic (\textbf{c}) and the fourth-harmonic voltage responses (\textbf{d}) of the TaIrTe$_4$ NLN device at 292.5 K, measured using each of the two heaters.}
    \label{fig:NLN_TIT_Data}
\end{figure}

Furthermore, we emphasize that any unintentional misalignment of the heater electrodes can generate a parasitic temperature gradient along the orthogonal direction. For instance, in the geometry employed to measure the NLN effect, this corresponds to a generation of a temperature gradient along $\hat{b}$ in the geometry illustrated in Fig. \ref{fig:NLN_TIT_Data}\textbf{a, b}, wherein $\nabla_a T$ corresponds to the desired temperature gradient. Such a parasitic temperature gradient can in turn produce a 2$\omega$ voltage response arising from the linear Seebeck effect, even in configurations where the heater electrodes are intended to generate a temperature gradient normal to the strip. If the parasitic contributions from both heaters are oriented along the same direction (for instance when both heater configurations share a common geometric imperfection - the same misalignment direction or the same thermal anchoring asymmetry), the resulting 4$\omega$ voltage response originating from the temperature dependence of the linear thermoelectric coefficients will contribute to the 4$\omega$ voltage response with the same sign, mimicking a contribution that can mistakenly be attributed to $|\nabla_a T|^2$. In the absence of a simultaneously measured 2$\omega$ voltage response, we emphasize that $V_{4\omega}$ as a function of the heater current alone does not allow one to unambiguously attribute the observed 4$\omega$ voltage response to the intrinsic nonlinear thermoelectric effects, even if the measured 4$\omega$ voltage response does not exhibit a sign change upon swapping the relative position of the active heater.\\

We illustrate this using the TaIrTe$_4$ device employed for the measurement of the NLN effect presented in the main text and the results shown in Fig.~\ref{fig:NLN_TIT_Data}. We observe a finite 2$\omega$ voltage response that does not change sign upon swapping the relative position of the active heater with respect to the flake, as shown in Fig.~\ref{fig:NLN_TIT_Data}\textbf{c}. The simultaneously measured 4$\omega$ voltage response also does not reverse sign upon swapping the position of the active heater, as shown in Fig.~\ref{fig:NLN_TIT_Data}\textbf{d}. In the absence of accounting for the temperature-induced nonlinearities originating from the temperature dependence of the linear thermoelectric coefficients, the entire 4$\omega$ voltage response could be incorrectly attributed to intrinsic nonlinear thermoelectric effects. Such an interpretation would lead to an overestimation of the intrinsic nonlinear thermoelectric coefficients and critically could result in assigning the 4$\omega$ voltage response that physically arises from the temperature-induced nonlinearities of the linear thermoelectric response to intrinsic nonlinear thermoelectric effects. It is thus crucial to also report the corresponding 2$\omega$ voltage response of the same voltage channel wherein the 4$\omega$ voltage response is measured.\\

We note that, for the configuration illustrated in Fig.~\ref{fig:NLN_TIT_Data}, an unintended temperature gradient along the strip, $\nabla_b T$, and the temperature gradient perpendicular to the strip, $\nabla_a T$, can be quantified through the linear thermoelectric responses, namely the Seebeck effect ($V_{2\omega}^{b} = -S_b  L \nabla_b T$) and the Nernst effect ($V_{2\omega}^{b} = \nu_{ba} B_c L\nabla_a T$), respectively. This enables an analysis analogous to that described above to isolate the intrinsic nonlinear thermoelectric response, namely the NLN effect. We further note that the extracted intrinsic nonlinear thermoelectric coefficient may contain contributions from both the NLN effect (from $\nabla_a T$) and the NLS response arising from the parasitic temperature gradient $\nabla_b T$. However, we note that $\nabla_a T$ about an order of magnitude larger than $\nabla_b T$. Furthermore, the coefficients associated with these two effects exhibit distinct temperature dependencies as shown in Fig.~3 of the main text, indicating that the parasitic NLS contribution is negligible. \\

Alternatively, the approach introduced in Ref.~[\onlinecite{SaitohPhase}], wherein both heaters are sourced simultaneously with a relative phase shift between the ac heater currents, equivalently provides a strategy to disentangle intrinsic nonlinear thermoelectric effects from temperature-induced nonlinearities. In this scheme, the use of two heaters with a controlled phase offset reduces the nonlinear thermoelectric response to the second-harmonic voltage response, while the linear thermoelectric voltage appears at the first-harmonic voltage response. Consequently, the dependence of the second-harmonic voltage response on the relative phase shift between the heater currents enables the effective measurement of intrinsic nonlinear thermoelectric effects from temperature-induced nonlinearities.\\

\subsection{Separation of the intrinsic nonlinear thermoelectric responses from the temperature-dependent heater resistance}

An additional pathway in which the linear thermoelectric effects can contribute to the measured nonlinear responses arises from the heater electrode itself. For an alternating current through the heater electrode, taking into account that the heater resistance depends on the local temperature, the voltage across it can be written as

\begin{equation*}
\label{eq:Pt_Heater_4o}
V_{\text{Heater}} = I(\omega) R(T)
= I(\omega) R_0 \bigl(1+\alpha \Delta T_\text{Heater} \bigr)
= I(\omega) R_0 \bigl[1+\beta I^2(\omega) \bigr],
\end{equation*}
where $R_0$ is the resistance of the heater electrode at temperature $T_0$, $\Delta T_\text{Heater} = T_\text{Heater} - T_0$ denotes the rise in the temperature of the heater due to Joule heating from an ac current through the heater, $\alpha$ (=$\frac{1}{R_0}\frac{\partial R}{\partial T}|_{T_0}$) characterizes the change in the heater resistance as a function of temperature, and $\beta$ parametrizes the current-induced temperature rise through Joule heating. Owing to the temperature dependence of the heater resistance, Joule heating in the heater produces an additional voltage response at $3\omega$ and thus also generates a temperature gradient at 4$\omega$ besides the temperature gradient at 2$\omega$. This contribution can be expressed as

\begin{equation*}
\nabla T|_{4\omega} \propto P_{\text{Heater}}|_{4\omega} = I(\omega)\, V(3\omega) = R_0 \alpha\, I^2 \Delta T_{\text{Heater}} = R_0 \beta\, I^4 .
\end{equation*}

As a consequence, the temperature gradient corresponding to $\nabla T|_{4\omega}$ (proportional to $\nabla T|_{2\omega}$) introduces a contribution from the linear thermoelectric response in the measured fourth-harmonic voltage response. This contribution is formally analogous to the effect arising from the temperature dependence of the linear thermoelectric coefficients, which can likewise generate a measurable fourth-order response, with the difference that the rise in temperature, namely $\Delta T$ is now that of the heater electrode itself unlike the rise in temperature of the flake in the previous case. In both cases, the observed signal scales with the fourth power of the driving ac current, but depends linearly on both (i) the temperature rise and (ii) the temperature gradient across the sample, the latter reversing sign when the heater position is inverted. 
\\

We emphasize that as a result of a large driving ac current in the heater, the temperature dependence of the heater resistance introduces a linear thermoelectric response proportional to $\nabla T|_{4\omega}$ in the measured fourth-order response. Both these contributions, namely the temperature dependence of the linear thermoelectric response and the temperature dependence of the heater resistance can be systematically separated from the intrinsic nonlinear thermoelectric effects using the analysis methodology described above as the directionality (sign) of these contributions to the measured nonlinear response depend on the direction of $\nabla_iT$. Furthermore, the rise in temperature of the heater can be obtained via resistance measurements of the heater as a function of the heater current and a low bias $R$-$T$ calibration as explained in detail in Sec.~\ref{Sec:Heater_Pt}. 


\subsection{Verification of contributions from the temperature-dependence of the linear thermoelectric effects} \label{Sec:Func_Form}

\begin{figure}[H]
    \centering
    \includegraphics[width=\textwidth]{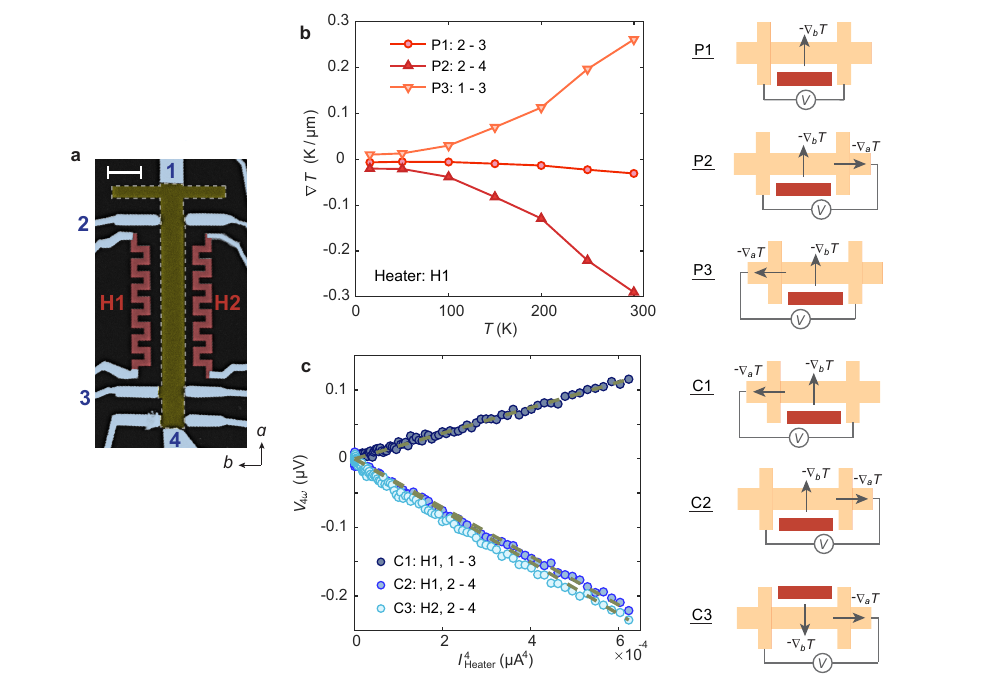}  
    \caption{\textbf{a}, False-coloured SEM image of the WTe$_2$ Hall bar device. WTe$_2$, Pt heaters, and Au contacts are shown in yellow, red, and light blue, respectively (Scale bar: $\SI{5} {\micro\meter}$).
    \textbf{b}, The longitudinal temperature gradient ($\nabla_a T$) as a function of temperature for an applied heater current of 5 mA, measured using three different pairs of voltage probes. The corresponding voltage-probes are numbered in panel \textbf{a}, and the schematics for each of the configuration are shown on the right side (P1, P2 and P3). The values of the temperature gradient were deduced from the linear thermoelectric effect measurements using the Seebeck coefficients determined from the longitudinal configuration. \textbf{c}, Fourth-harmonic voltage response as a function of the fourth power of the heater current at 15 K, measured using the three different configurations schematically shown on the right side (C1, C2 and C3). The dashed gray lines are the linear fitting lines.}
    \label{fig:trans_gradT}
\end{figure}

We further confirm the contribution of the nonlinearities arising from the temperature dependence of the linear thermoelectric effects in the measured 4$\omega$ voltage response, as well as from parasitic temperature gradients utilizing WTe\textsubscript{2} Hall bar devices. The Hall bar devices were fabricated on Si/SiO$_2$ substrates, and the platinum heater electrodes are placed between the transverse arms of the Hall bar geometry as shown in Fig.~\ref{fig:trans_gradT}\bf{a}\normalfont. As a result, the length of the channel where the voltage drop is measured is longer than the length of the heater electrode, which inevitably leads to a considerable longitudinal temperature gradient in addition to the desired transverse temperature gradient within the WTe$_2$ flake. By using different pairs of voltage probes, we confirm that the longitudinal temperature gradient reverses upon changing the lateral position of the voltage probes with respect to the heater electrode, which is inferred from the sign of the linear thermoelectric voltage response (Fig.~\ref{fig:trans_gradT}\bf{b}\normalfont).\\

The $4\omega$ voltage response measured in the Hall bar geometry was investigated by deliberately introducing an asymmetry in the voltage detection channel as shown in Fig.~\ref{fig:trans_gradT}\textbf{c}.
When the longitudinal temperature gradient (along the channel, namely $\nabla_a T$) is flipped between the configurations C1 and C2 (while the transverse temperature gradient is kept the same by using the same heater), the $4\omega$ voltage response exhibits a clear sign reversal.
In contrast, when the transverse temperature gradient normal to the Hall channel is inverted, namely when $\nabla_bT$ is reversed by interchanging the heater position between configurations C2 and C3 (note that the voltage probes are chosen such that $\nabla_a$T does not change sign), we observe that the sign of the fourth-harmonic voltage response remains unchanged, with only its magnitude being slightly modified. \\

These results demonstrate that the sign (directionality) of the observed $4\omega$ voltage response predominantly originates from a contribution that is uncorrelated with the transverse temperature gradient ($\nabla_{b}T$), but instead depends on the longitudinal temperature gradient ($\nabla_{a}T$). In particular, the sign of the $4\omega$ response is governed by the direction of $\nabla_{a}T$, consistent with the term of the form given in Eq.~\ref{NLE_Linear} as discussed previously. The difference in amplitude observed between configurations C2 and C3 in Fig.~\ref{fig:trans_gradT}\textbf{c} can be attributed to discrepancies in the temperature gradients generated by the two heaters, as well as to the intrinsic nonlinear thermoelectric contribution (specifically $S_{aab}^{(2)}$) of WTe\textsubscript{2}.
\subsection{Fourth-harmonic voltage responses arising from the nonlinearities associated with the temperature dependence of the linear thermoelectric effects and the heater resistance}

The contributions of the nonlinearities arising from the temperature dependence of the linear thermoelectric effects and the heater resistance in the fourth-harmonic voltage response were extracted by anti-symmetrizing the measured fourth-harmonic ($4\omega$) voltage responses as a function of heater current while swapping the active heater, following the procedure explained above. By combining the linear thermoelectric response obtained from the second-harmonic voltage measurements with the heater temperature rise inferred from resistance–temperature calibration, we separate the nonlinear contributions arising from the temperature dependence of the linear thermoelectric coefficients and the heater resistance. 

\begin{figure}[H]
    \centering
    \makebox[\textwidth][c]{%
    \includegraphics[width=1.0\textwidth]{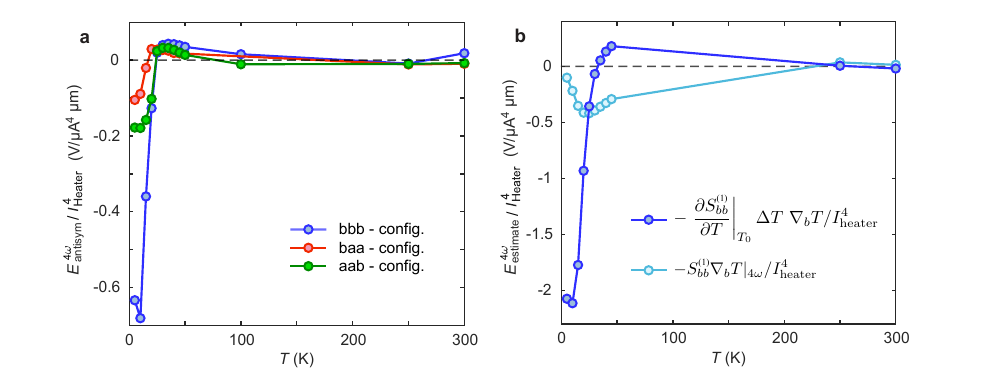}   
    }
    \caption{\textbf{a}, The contribution from the temperature dependence of the linear thermoelectric effect obtained from the anti-symmetrized fourth-harmonic voltage response, normalized by the heater current for the WTe\textsubscript{2} devices with three different geometries presented in the main text. bbb, baa, and, aab - config. mean the sample geometries used to measure $S_{bbb}^{(2)}$, $S_{baa}^{(2)}$, and $S_{aab}^{(2)}$ respectively. Error bars are not shown due to their small magnitude. \textbf{b}, The estimated  contributions from the nonlinearity arising due to the temperature dependence of the linear thermoelectric effect and the heater resitance that produces a linear temperature gradient at 4$\omega$ for the NLS sample. Both components are calculated using the measured linear Seebeck coefficient as explained in the text.}
    \label{fig:dSdT}
\end{figure}

The extracted temperature-induced nonlinear contributions of WTe\textsubscript{2} are shown in Fig.~\ref{fig:dSdT}\textbf{a}. We note that the extracted ${\partial S_{ij}^{(1)}}/{\partial T}$ correspond to ${\partial S_{bb}^{(1)}}/{\partial T}$, ${\partial S_{ba}^{(1)}}/{\partial T}$ and a combination of both ${\partial S_{ab}^{(1)}}/{\partial T}$ and ${\partial S_{aa}^{(1)}}/{\partial T}$ for the sample geometries used to measure $S_{bbb}^{(2)}$, $S_{baa}^{(2)}$ and $S_{aab}^{(2)}$ respectively. The observed trend extracted for ${\partial S_{ij}^{(1)}}/{\partial T}$ from the procedure detailed above is qualitatively consistent with the previous reports of ${\partial S_{ii}^{(1)}}/{\partial T}$ reported for bulk WTe\textsubscript{2} in Ref.~\citen{wu2015temperature}.\\

By comparing the extracted nonlinear contributions associated with the temperature dependence of the linear thermoelectric coefficient and the heater resistance as shown in Fig.~\ref{fig:dSdT}\textbf{b}, we find that although the temperature dependence of the heater resistance yields a noticeable contribution to the anti-symmetrized fourth-harmonic voltage response, the dominant contribution arises from the temperature dependence of the linear thermoelectric response.\\

We further attribute the discrepancy in the magnitude between the measured anti-symmetrized fourth-harmonic voltage response and the estimated contributions (from the measured linear thermoelectric responses) to an overestimation of both $\Delta T$ and $\nabla T$ in the finite-element simulations (see Sec.~\ref{Sec:COMSOL}). A likely origin of this overestimation is the presence of Ti/Au contact electrodes, which act as heat sinks in the actual device geometry but are not fully captured in the model. Consequently, we emphasize that the inferred nonlinear thermoelectric coefficients from the symmetrized fourth-harmonic voltage response should be regarded as lower bounds.
 
\section{Absence of detectable symmetry-forbidden nonlinear thermoelectric responses}

\begin{figure}[H]
    \centering
    \makebox[\textwidth][c]{%
    \includegraphics[width=1.0\textwidth]{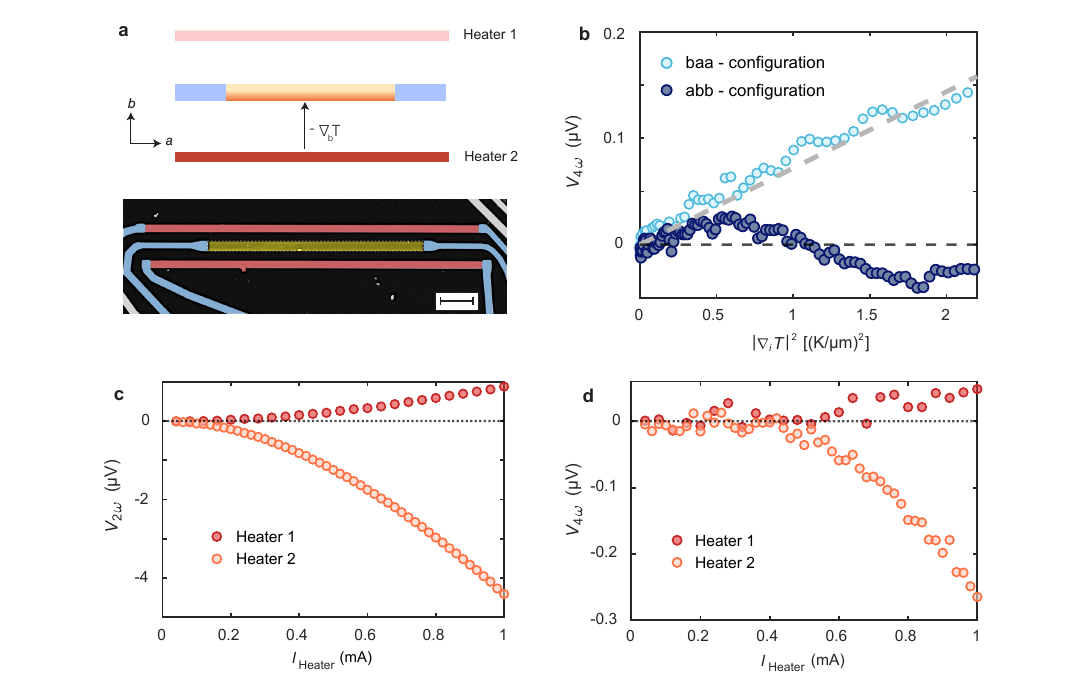}   
    }
    \caption{\textbf{a}, Schematic illustration (top) and the corresponding false-coloured SEM images (bottom) of the devices designed to measure the NLS effect for the symmetry-forbidden device configuration. In the SEM images, WTe\textsubscript{2}, Pt heaters, and Au contacts are shown in yellow, red, and light blue, respectively (Scale bar: $\SI{10} {\micro\meter}$). \textbf{b}, Fourth-harmonic voltage response, symmetrized with respect to the heater switching, as a function of the square of the temperature gradient at 45 K of the device shown in panel \textbf{a}. The corresponding data for the device fabricated to probe the symmetry-allowed $S^{(2)}_{baa}$ are also shown for comparison. \textbf{c}-\textbf{d}, Raw data of the second-harmonic (\textbf{c}) and the fourth-harmonic voltage (\textbf{d}) responses corresponding to panel \textbf{b}, measured by generating a temperature gradient by driving an ac current in the two different heater electrodes.}
    \label{fig:forbidden_NLN}
\end{figure}

To establish that the observed nonlinear thermoelectric responses originate from the reduced crystalline symmetry as highlighted in Sec.~\ref{Sec:Symm}, we designed and fabricated additional device geometries to test for the nonlinear thermoelectric components that are forbidden by symmetry, as specified in Eq.~\ref{Reduced_Matrix_form}. The devices fabricated to probe the nonlinear thermoelectric response corresponding to $S^{(2)}_{abb}$ is shown in Fig.~\ref{fig:forbidden_NLN}. The symmetrized $4\omega$ voltage response for the symmetry-forbidden NLN configuration is presented in Fig.~\ref{fig:forbidden_NLN}\textbf{b}. Notably, we do not observe the characteristic $|\nabla_iT|^{2}$ scaling in the abb configuration, in contrast to the clear response observed in the baa configuration. The absence of such a 4$\omega$ voltage response in the configuration of $S^{(2)}_{abb}$ confirms that the observed nonlinear response originates from the reduced crystal symmetry as highlighted in Sec.~\ref{Sec:Symm}.\\

Additionally, from the measured raw second- and fourth-harmonic  voltage responses shown in Figs.~\ref{fig:forbidden_NLN}\textbf{c} and \textbf{d}, respectively, it can be seen that the voltage responses scale consistently with the heater current for both the second- and fourth-harmonic voltage responses. We emphasize that in this case, the nonlinear thermoelectric voltage response observed can be attributed completely to the temperature dependence of the second order response (or the temperature dependence of the heater resistance).\\

We emphasize that a residual longitudinal temperature gradient along the channel can generate an intrinsic contribution in this geometry via the $S^{(2)}_{aab}$ tensor component. Additionally, as discussed in Sec.~\ref{Sec:Off_Axis_NL}, a source of the measured 4$\omega$ voltage response could be a slight misalignment of the voltage detection channel with respect to the crystallographic \textit{a}-axis that causes an introduction of the symmetry-allowed nonlinear thermoelectric responses.\\

\begin{figure}[H]
    \centering
    \makebox[\textwidth][c]{%
    \includegraphics[width=1.0\textwidth]{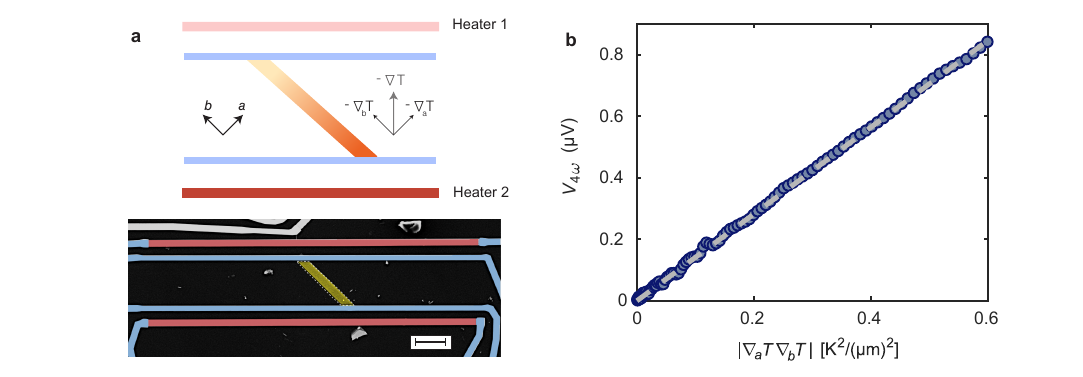}   
    }
    \caption{\textbf{a}, Schematic illustration (top) and the corresponding false-coloured SEM images (bottom) of the devices designed to measure the NLMT effect for the symmetry-forbidden device orientation. In the SEM images, WTe\textsubscript{2}, Pt heaters, and Au contacts are shown in yellow, red, and light blue, respectively (Scale bar: $\SI{10} {\micro\meter}$). \textbf{b}, Fourth-harmonic voltage response, symmetrized with respect to the heater switching, as a function of the square of the temperature gradient at 45 K of the device shown in panel \textbf{a}. 
    }
    \label{fig:forbidden_NLMT}
\end{figure}

Although we confirm the absence of $S^{(2)}_{abb}$ using our device geometry and the measurement scheme, we address that it is difficult to apply the same scheme for confirming the case of the NLMT response $S_{bab}^{(2)}$ (Fig.~\ref{fig:forbidden_NLMT}). This is because, although the contribution from $S_{bab}^{(2)}$ is expected to be zero, the intrinsic nonlinear thermoelectric contribution of the NLS and the NLN, namely $S^{(2)}_{bbb}$ and $S_{baa}^{(2)}$, affects this device geometry as follows:

\begin{equation}
\label{Eq:Forbidden_Sbab}
    \begin{aligned}    
    V_{4\omega} & = + [S_{bbb}^{(2)} (\nabla_b T)^2 + S_{baa}^{(2)} (\nabla_a T)^2] L \quad.
    \end{aligned}
\end{equation}

We observe a pronounced 4$\omega$ voltage response in the configuration of $S^{(2)}_{bab}$ as shown in Fig.~\ref{fig:forbidden_NLMT}\textbf{b}, likely arising due to the the combined effects of both the NLS and NLN effect as given in Eq.~\ref{Eq:Forbidden_Sbab}. 
\newpage
\section{Excluding the influence of the capacitive coupling between the heater electrode and the voltage probe} \label{Sec:Capacitive}

\begin{figure}[H]
    \centering
    \makebox[\textwidth][c]{%
    \includegraphics[width=1\textwidth]{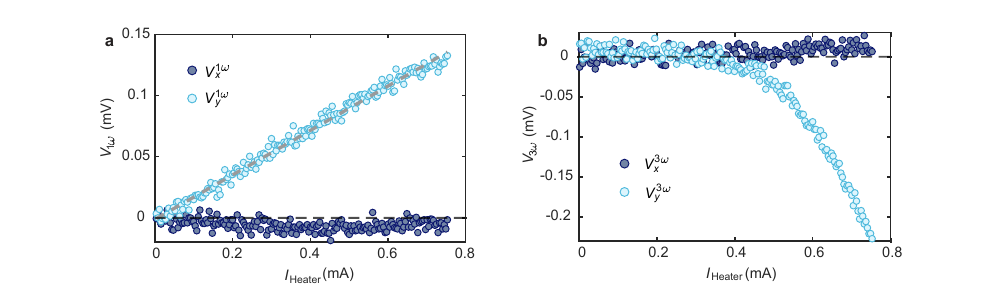}
    }
    \caption{\textbf{a}-\textbf{b}, The first-harmonic (\textbf{a}) and third-harmonic (\textbf{b}) voltage for the WTe\textsubscript{2} NLS configuration device as a function of the heater current at 5 K ($f$ = 17.777 Hz).}
    \label{fig:capacitive}
\end{figure}

The first- and third-harmonic voltage responses of WTe\textsubscript{2} as a function of the heater bias current are shown in Fig.~\ref{fig:capacitive}\textbf{a} and \textbf{b}, respectively. While a first-harmonic voltage response is detected, the coupling between the heater electrode and the voltage probe is primarily capacitive, whose magnitude furthermore scales proportional to $j/(C\omega)$ upon changing the frequency of the ac current. This capacitive coupling arises from both the coupling across the measurement cables and the direct coupling between conductors on the device itself.
\\

The appearance of the $3\omega$ voltage response signal can be attributed to the Joule heating in the Pt heater electrode. As detailed before, the voltage drop across the heater can be written as

\begin{equation} \label{eq:RT_Pt}
\begin{aligned}
V_{\text{Heater}} &= I(\omega) R(T) \\
&= I(\omega) R_0\bigl(1+\alpha \Delta T\bigr), \\
\end{aligned}
\end{equation}
showing that the temperature-dependent resistance introduces both first- and third-harmonic voltage responses across the heater electrode. These voltage responses then capacitively coupled to the detector electrode, produces correspondingly a first- and a third-harmonic voltage response in the out-of-phase component of the measurements.

We note that any parasitic leakage of odd-harmonic signals, namely $3\omega$, $5\omega$, and higher orders can be ruled out as the origin of the measured even-harmonic responses, specifically the $2\omega$ and $4\omega$ voltage responses that are attributed to the linear and nonlinear thermoelectric effects, respectively. Since the odd-harmonic components arise from capacitive coupling between the heater and the voltage probe, which scales as $j/(C\omega)$, measurements performed as a function of the frequency of the heater current can exclude capacitive coupling as the source of the observed even-harmonic responses as presented in Sec.~\ref{sec:Freq_Dep}. 


\section{Frequency dependence and exclusion of dc offset}
\label{sec:Freq_Dep}

\begin{figure}[H]
    \centering
    \includegraphics[width=\textwidth]{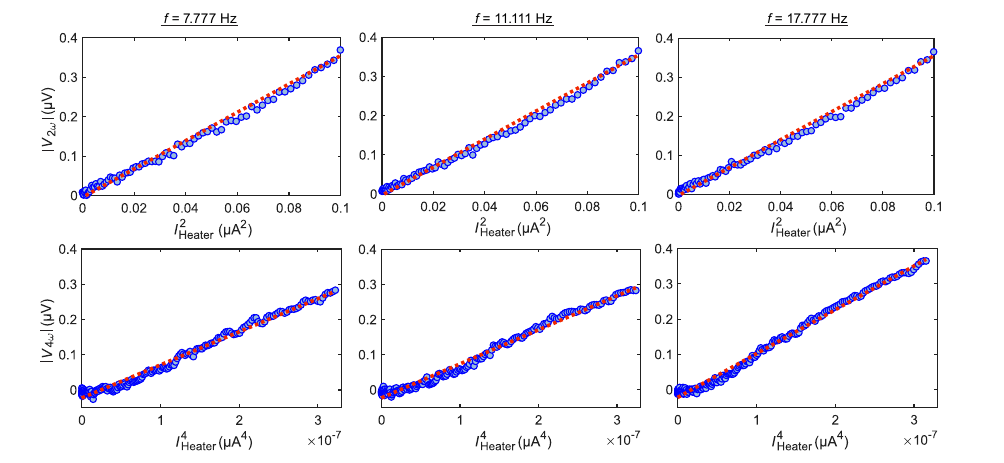} 
    \caption{Second- (upper panels) and fourth-harmonic (lower panels) voltage responses as a function of applied heater current at 5 K, measured at frequencies of the 7.777, 11.111, and 17.777 Hz. The sample measured is the WTe$_2$ NLS configuration device on fused silica.}
    \label{fig:freq_fusedSiO2}
\end{figure}

\begin{figure}[H]
    \centering
    \includegraphics[width=\textwidth]{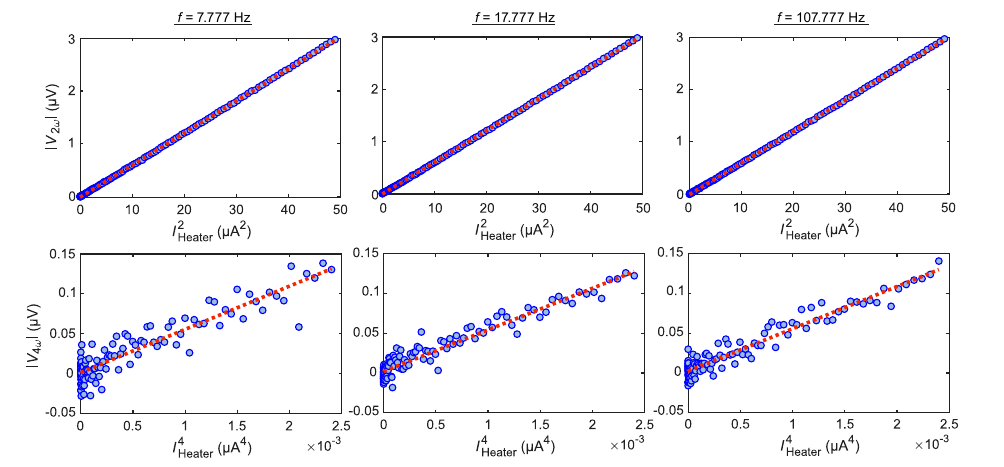}
    \caption{Second- (upper panels) and fourth-harmonic (lower panels) voltages as a function of applied heater current at 15 K, measured at frequencies of 7.777, 17.777, and 107.777 Hz. The sample measured is the the WTe$_2$ NLS configuration device on Si/SiO$_2$ substrate.}
    \label{fig:freq_largePt}
\end{figure}

The frequency dependence of the second and the fourth-harmonic voltage responses measured for WTe$_2$ on both fused silica and Si/SiO$_2$ substrates are shown in Fig.~\ref{fig:freq_fusedSiO2} and Fig.~\ref{fig:freq_largePt} respectively. We note that since the voltage responses are independent of the driving frequency of the heater current, the capacitive coupling between the heater electrode and the voltage detection channel as described in Sec.~\ref{Sec:Capacitive} does not contribute to the measured thermoelectric voltage responses as we expect any parasitic contribution from the capacitive coupling to scale as $j/(C\omega)$. Furthermore, the frequency independence of the measured thermoelectric voltage responses indicate that the driving frequency ($f$) satisfies $f < 1/\tau_{\mathrm{th}}$, where $\tau_{\mathrm{th}}$ is the thermal time constant of the substrate. Note that  $\mu A^4$ in Fig.~\ref{fig:freq_fusedSiO2} and Fig.~\ref{fig:freq_largePt} is used to denote 10$^{-6}A^4$. \\

\subsection{Frquency dependence of the thermoelectric responses}

\subsection{Exclusion of a dc offset in the current source}

\begin{figure}[H]
   \centering
        \makebox[\textwidth][c]{%
    \includegraphics[width=0.9\textwidth]{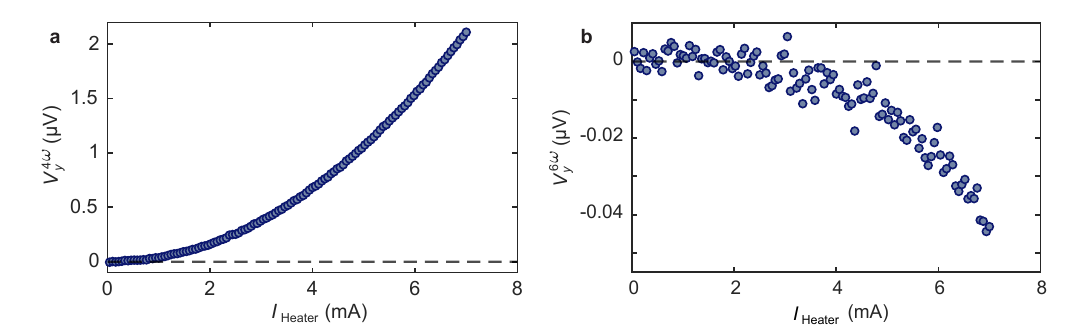}} 
    \caption{\textbf{a}-\textbf{b}, Fourth- (\textbf{a}) and the sixth-harmonic (\textbf{b}) voltage responses as a function of applied heater current at 15 K ($f$ = 17.777 Hz) of the WTe$_2$ device used for the measurement of the anisotropy in the Seebeck coefficients on Si/SiO$_2$ substrate.}
    \label{fig:6Omega}
\end{figure}

We note that the presence of a dc offset in the ac current sourced can lead to parasitic mixing of higher-order nonlinear responses into the lower harmonic voltage responses. In particular, a sixth-order response in the heater current can contribute to the fourth-harmonic voltage response. This can be seen by expanding the sixth-order voltage response as

\begin{equation}
\label{Eq:Six_in_Four}
\begin{aligned}
V_6 &= R_6\left[I(t)\right]^6 = R_6\left[I_{dc}+\sqrt{2}I_0\sin(\omega t)\right]^6 \\
V_{6\rightarrow4}&= \binom{6}{2} I_{dc}^{2}\left[\sqrt{2}I_0\sin(\omega t)\right]^{4} R_6 ,
\end{aligned}
\end{equation}
where $\binom{6}{2}$ is the binomial coefficient. The presence of a dc component thus generates terms proportional to $\sin^4(\omega t)$, that contributes to a fourth-harmonic voltage response. Consequently, the sixth-order response partially contributes to the measured fourth-harmonic voltage. The fourth- and sixth-harmonic voltage responses measured simultaneously are shown in Fig.~\ref{fig:6Omega}.

From the amplitudes of the measured fourth- and sixth-harmonic voltage responses, and from Eq.~\ref{Eq:Lockin} and Eq.~\ref{Eq:Six_in_Four}, we estimate that a dc offset of approximately $\sim$ 8.5 mA in the heater current would be required for the measured fourth-harmonic signal to arise entirely from the sixth-order response. This value is unrealistic, as the maximum rms amplitude of the heater current is only 7 mA. Furthermore, a direct estimate based on Eq.~\ref{Eq:Lockin} shows that the sixth-order contribution to the fourth-harmonic voltage amounts to only about $4\%$ of the fourth-order contribution, validating that the fourth-harmonic voltage response measured is dominated by the fourth order responses. 


\section{Nonlinear Hall effect}

\begin{figure}[H]
    \centering
        \makebox[\textwidth][c]{%
    \includegraphics[width=1.0\textwidth]{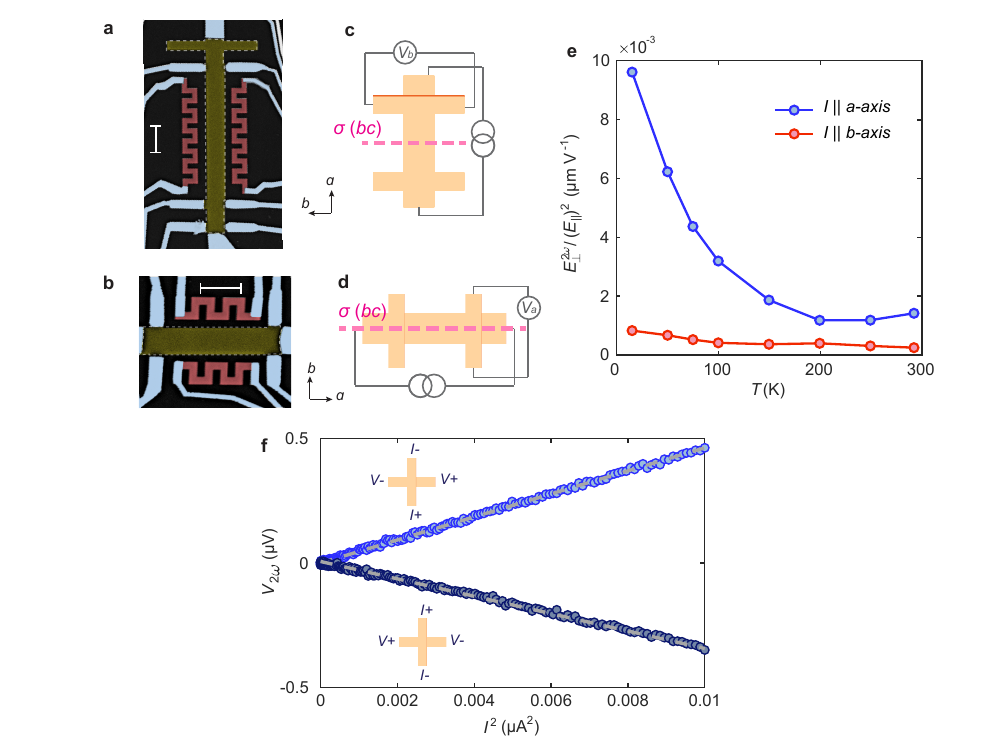}   
    }
    \caption{\textbf{a}-\textbf{b}, False-coloured SEM image of the WTe$_2$ Hall bar devices. The WTe$_2$ is oriented such that the Hall channel wherin the charge current flows is oriented along the crystallographic \textit{a}-(\textbf{a}) and \textit{b}-axes (\textbf{b}). WTe$_2$, Pt heaters, and Au contacts are shown in yellow, red, and light blue, respectively (Scale bar: $\SI{5} {\micro\meter}$). \textbf{c}-\textbf{d}, Schematic illustrations of the device geometry and the measurement scheme for the nonlinear Hall measurements corresponding to the SEM images in panels \textbf{a} and \textbf{b}.
    \textbf{e}, Temperature dependence of the nonlinear Hall response measured along the two different crystallographic axis orientations using the devices shown in panels \textbf{a} and \textbf{b}. Error bars are smaller than the data points.
    \textbf{f}, 
    The quadratic dependence of the nonlinear Hall voltage on the bias current at 15 K, measured under two different configurations wherein both the direction of the charge current and the polarity of the voltage detection are simultaneously swapped. The current direction is along the crystallographic \textit{a}-axis. Dashed gray lines indicate the corresponding linear fits.}
    \label{fig:NLH}
\end{figure}

To better understand the scaling between the measured NLN effect and the nonlinear Hall (NLH) effect as a function of temperature, we utilize the Hall-bar device (Fig.~\ref{fig:NLH}\textbf{a}) of WTe$_2$. The NLN effect is the thermal equivalent to the NLH effect under the constraints of the crystal symmetry and a formal phenomenological treatment of both effects follow along identical lines to that presented in Sec.~\ref{Sec:Symm}. The symmetry constraints of the NLH effect in WTe$_2$ has been verified and have been reported in Ref.~[\onlinecite{ma2019observationSI}] and  Ref.~[\onlinecite{kang2019nonlinear}]. The NLH measurements similar to the NLN measurements were measured in unencapsulated WTe$_2$ flakes. \\

As shown in Fig. \ref{fig:NLH}\textbf{e}, a clear NLH effect is observed when the charge current is oriented along the crystallographic $a$-axis and the transverse voltage is measured along the $b$-axis. In contrast, when the charge current is oriented along the $b$-axis, the second harmonic voltage response remains approximately by an order of magnitude smaller in comparison to when the charge current is sourced along the  crystallographic $a$-axis, consistent with the symmetry constraints.
The temperature dependence follows the trend reported previously in few-layer WTe$_2$~\cite{kang2019nonlinear}. The residual transverse voltage response that is observed when the charge current is along the crystallographic \textit{b}-axis can be attributed to the small misalignment of the voltage detection channel from the high symmetry crystallographic axis.\\ 

We confirm that the NLH voltage response change sign when both the direction of the charge current and the polarity of voltage detection are reversed simultaneously (Fig.~\ref{fig:NLH}\bf{b}\normalfont). This indicates that the detected voltage response remains unaffected by the reversal of the direction of the charge current, confirming that it originates from the nonlinear second-order dependence on the charge current.

\section{Scaling law for the nonlinear thermoelectric responses}

To understand the microscopic origin of the nonlinear thermoelectric responses in WTe$_2$ and TaIrTe$_4$, we examine the temperature dependence of the nonlinear thermoelectric coefficients as a function of the corresponding longitudinal conductivity and the linear thermoelectric coefficients. Similar scaling analysis has been widely employed in studies of nonlinear electrical transport, particularly for the NLH effect~\cite{bandyopadhyay2024non,du2019disorder}, where such scaling analysis provides a powerful framework in disentangling the underlying microscopic mechanisms governing the nonlinear responses. Within semiclassical Boltzmann transport theory, the longitudinal electrical conductivity is directly proportional to the transport relaxation time ($\tau$), $\sigma_{ii}\propto\tau_i$, and therefore serves as an experimental proxy for the carrier scattering. Analyzing the scaling of the NLH conductivity as a function of the longitudinal conductivity has enabled the separation of skew-scattering, side-jump, and intrinsic contributions, each characterized by a distinct dependence on the longitudinal conductivity and, consequently, on the scattering time \cite{kang2019nonlinear,kumar2021room}.\\

An analogous scaling framework has recently been extended to the nonlinear Nernst effect in ABA trilayer graphene ~\cite{liu2025nonlinear}, where the nonlinear Nernst coefficient was reported to approximately follow the relation $S_{yxx}^{(2)}\propto[S_{xx}^{(1)}\sigma_{xx}]^2$. This behavior was interpreted as evidence for a dominant skew-scattering contribution and was argued to be consistent with the crystal symmetry of ABA trilayer graphene, where the $C_3$ rotational symmetry suppresses the Berry-curvature-dipole contribution. More generally, this approach has been extended theoretically to the nonlinear thermoelectric conductivity tensor, providing a route to identify the microscopic mechanisms governing nonlinear thermoelectric transport \cite{varshney2026asymmetric}.\\

\begin{figure}[H]
    \centering
        \makebox[\textwidth][c]{%
    \includegraphics[width=1.0\textwidth]{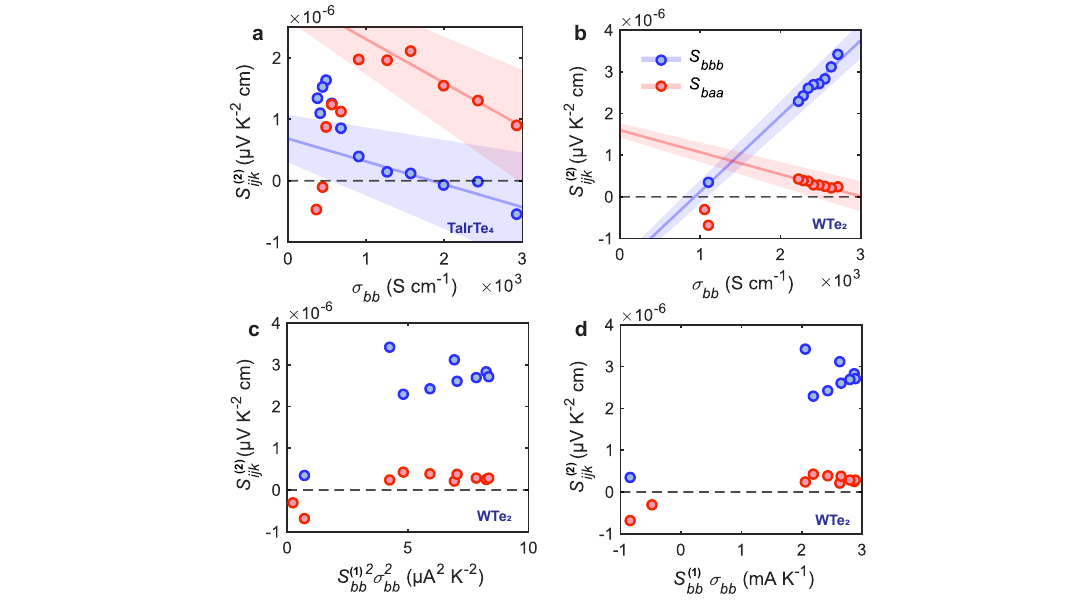}   
    }
    \caption{Comparison of different scaling relations for the NLS and NLN coefficients. \textbf{a}-\textbf{b}, The NLS and NLN coefficients in TaIrTe$_4$(\textbf{b}) and WTe$_2$(\textbf{b}) as a function of electrical conductivity along $b$-axis ($\sigma_{bb}$). Linear fits to the data below 55 K are shown for panels, with the shaded regions representing the standard deviation of the fits. The electrical conductivity data for thin TaIrTe$_4$ were extracted from Ref.~\citen{Gonio_TIT}. \textbf{c}-\textbf{d}, The NLS and NLN coefficients in WTe$_2$ as a function of $[S_{bb}^{(1)}\sigma_{bb}]^2$ (\textbf{c}), and $S_{bb}^{(1)}\sigma_{bb}$ (\textbf{d}). }
    \label{fig:scaling}
\end{figure}

However, we find that the nonlinear thermoelectric coefficients in WTe$_2$ and TaIrTe$_4$ are more accurately described by the phenomenological expression

\begin{equation}
S_{ijk}^{(2)} = A\sigma_{ii}^{2} + C.
\label{scaling_eq}
\end{equation}
Within a disorder-mediated nonlinear transport picture, the quadratic term is generally associated with skew-scattering process, whereas the constant term captures contributions from intrinsic mechanisms. The relative magnitudes of $A$ and $C$ therefore provide insight into the dominant microscopic processes governing the nonlinear thermoelectric response.\\

The improved description obtained utilizing $\sigma_{ii}$ rather than $S_{ii}^{(1)}\sigma_{ii}$ (Figs.~\ref{fig:scaling}\textbf{c} and \textbf{d}) likely reflects the breakdown of the assumptions underlying the simple Mott-Boltzmann picture over the extended temperature range considered here. In the low-temperature limit, the Seebeck coefficient is expected to be only weakly dependent on the relaxation time, such that $S_{ii}^{(1)}\sigma_{ii}$ serves as a direct proxy for the thermoelectric conductivity. However, both WTe$_2$ and TaIrTe$_4$ possess complex multiband electronic structures, and the temperature range studied spans a large window wherein phonon scattering and chemical-potential shifts (such as Lifshitz transition in WTe$_2$, see Ref. \citen{wu2015temperature}) become increasingly important. As a result, $S_{ii}^{(1)}(T)$ might acquire significant temperature dependence that is not directly linked to the scattering processes governing the nonlinear response. This interpretation is consistent with the temperature dependence of the linear thermoelectric coefficients and the electrical conductivity discussed in Sec.~\ref{Linear_Thermo} and Sec.~\ref{Sec:Anisotropic_Response}, indicating that the simple Mott-Boltzmann assumptions are not valid across the full experimental temperature range.\\

Owing to the limited number of data points, we performed two-parameter fitting using either $A\sigma_{ii}^{2} + C$ or $B\sigma_{ii} + C$, rather than a three parameter fitting based on with $A\sigma_{ii}^{2} + B\sigma_{ii} + C$. In addition to the quadratic scaling relation shown in Fig.~3 in the main text, we note that the linear scaling with electrical conductivity (Fig.~\ref{fig:scaling}\textbf{a} and \textbf{b}) also provides a reasonable description of the experimental data. However, the linear-in-$\sigma_{ii}$ scaling of the NLS response exhibits a substantial constant term, namely $C$. Recent theoretical work has shown that intrinsic mechanisms that scale linearly with the scattering time ($\tau$), cannot contribute to the NLS effect \cite{varshney2026asymmetric}. We therefore model the nonlinear thermoelectric responses utilizing the scaling of the form $A\sigma_{ii}^{2} + C$, excluding the linear-in-$\sigma_{ii}$ term.
Further detailed studies are required to disentangle disorder-mediated contributions from material-specific electronic structure effects. In particular, nonlinear thermoelectric measurements on gated heterostructures may provide a promising route towards identifying the dominant mechanisms governing the nonlinear thermoelectric response.



\section{Finite-element simulation of the temperature distribution} \label{Sec:COMSOL}\

\subsection{Simulation for devices with fused silica substrate} \label{Sec:Heater_Pt}

\begin{figure}[htb]
    \centering
    \makebox[\textwidth][c]{%
    \includegraphics[width=1.05\textwidth]{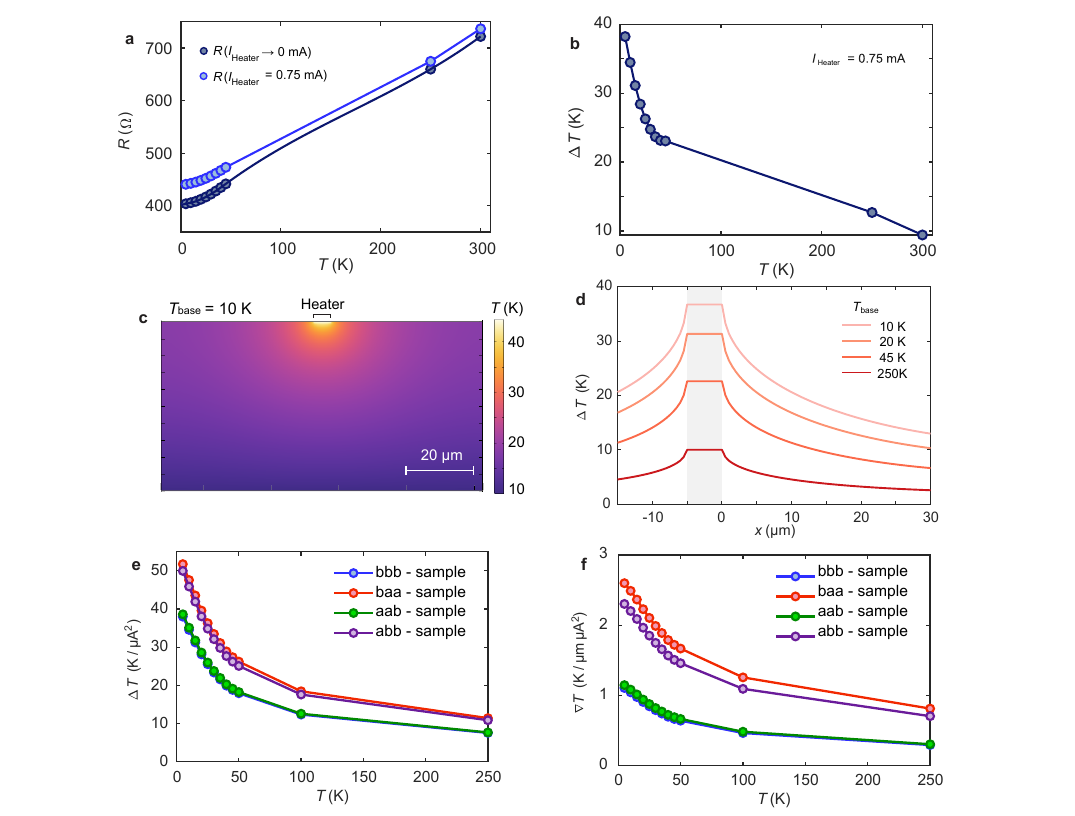}   
    }
    \caption{\textbf{a}, The resistance of the Pt heater electrode as a function of the base temperature for the zero-current limit and for an applied current of 0.75 mA for the WTe$_2$ NLN configuration device. The resistance corresponding to the zero-current limit was extracted by fitting the current dependence of heater resistance as described in the text. To further improve the accuracy of the estimation of the temperature rise of the heater, the temperature dependence of the resistance corresponding to the zero-current limit was interpolated using a spline function. \textbf{b}, Temperature rise (defined as $\Delta$ T = T$_\mathrm{heater}$ - T$_\mathrm{base}$) for a heater current of $I_\mathrm{Heater}$ = 0.75 mA estimated from the change in resistance shown in panel \textbf{a}. \textbf{c}, The cross-sectional simulation model and the resulting temperature distribution for a base temperature ($T_\mathrm{base}$) of 10 K. \textbf{d}, The simulated temperature profile along the upper edge of the fused silica substrate for various base temperatures. The gray shaded range corresponds to the position of Pt heater electrode with a fixed temperature. \textbf{e}, The average temperature rise over the regions corresponding to the WTe$_2$ strip positions in the devices. \textbf{f}, The temperature gradient normal to the Pt strips for each of the device configurations estimated from the WTe$_2$ strip positions in the actual devices.
}
    \label{fig:simu_fusedSiO2}
\end{figure}

The temperature gradient in the WTe\textsubscript{2} and TaIrTe\textsubscript{4} flakes were estimated using finite-element simulations using the COMSOL software~\cite{COMSOL}. 
The temperature distribution was simulated based on the experimentally confirmed temperature increase of the Pt heaters.\\

First, the dependence of the platinum heater resistance on the applied bias current was measured simultaneously along with the thermoelectric measurements, and was confirmed to follow the $R-T$ scaling described in Eq.~\ref{eq:RT_Pt}. The resistance of one of the Pt heaters as a function of the base temperature, measured for both the low-current limit and for an applied bias current of $0.75~\mathrm{mA}$, are shown in Fig.~\ref{fig:simu_fusedSiO2}\textbf{a}. The temperature increase corresponding to a given bias current was then extracted from the temperature dependence of the heater resistance (Fig.~\ref{fig:simu_fusedSiO2}\textbf{b}). Although slight differences in heating efficiency may be expected for the different heaters, no measurable difference in the temperature rise of the Pt heaters as a function of the heater current was observed within the experimental sensitivity across the two heater electrodes. We further confirm that all the heater electrodes have similar widths and thicknesses (as confirmed by Scanning Electron Microscopy and Atomic Force Microscopy respectively). \\

Using the estimated heater temperature rise as an input, finite-element simulations were carried out with the COMSOL Multiphysics software~\cite{COMSOL}.
We constructed a cross-sectional model that represents the actual device geometry. The simulation geometry along with a representative result is presented as a temperature map in Fig.~\ref{fig:simu_fusedSiO2}\textbf{c}. A rectangular region representing the fused silica substrate was defined, and the temperature at a specific edge region of the amorphous SiO$_x$ was fixed to the heater temperature estimated from the procedure highlighted above. This region was considered as the interfacial region between the heater and the substrate. We note that in this procedure we assume the temperature rise in the heater electrode corresponds to the interface temperature between the heater and the substrate, which does not account for the interfacial thermal resistance, namely the Kapitza resistance or the temperature dependence of the Kapitza resistance. The bottom edge of the amorphous SiO$_x$ was fixed at the base temperature of the measurement as a boundary condition. It was verified that the temperature distribution in the region of interest remained essentially unchanged when the size of the amorphous SiO$_x$ domain was increased. We further include the Au leads on selected regions on top of the amorphous SiO$_x$ domain, corresponding to the Au contacts in the actual devices. The dimensions of the Au regions were chosen such that the simulated temperature rise was in agreement with the experimentally observed temperature increase. We do not include WTe\textsubscript{2} or TaIrTe\textsubscript{4} as separate elements in the model, as we note that its presence has a negligible effect on the temperature distribution in the \textit{xy}-plane (see Sec.~\ref{Sec:Grad_OOP}). The simulated temperature gradient across a selected region along the upper edge of the amorphous SiO$_x$ is shown for several temperatures in Fig.~\ref{fig:simu_fusedSiO2}\textbf{d}. \\

The temperature rise and the temperature gradient in the sample flakes were determined by considering their position and size relative to the Pt heater in the actual devices that was estimated from scanning electron micrographs. 
The derived  temperature rise and the temperature gradient for WTe\textsubscript{2} samples are shown in Fig.~\ref{fig:simu_fusedSiO2}\textbf{e} and \textbf{f} (we note that the $y$-axis units of K/$\mu$A$^{2}$ corresponds to K/(mA)$^{2}$). 

\subsection{Simulation for devices with Si/SiO$_2$ substrates}
\label{Sec:simu_Si}

\begin{figure}[htb]
    \centering
     \makebox[\textwidth][c]{%
    \includegraphics[width=1.0\textwidth]{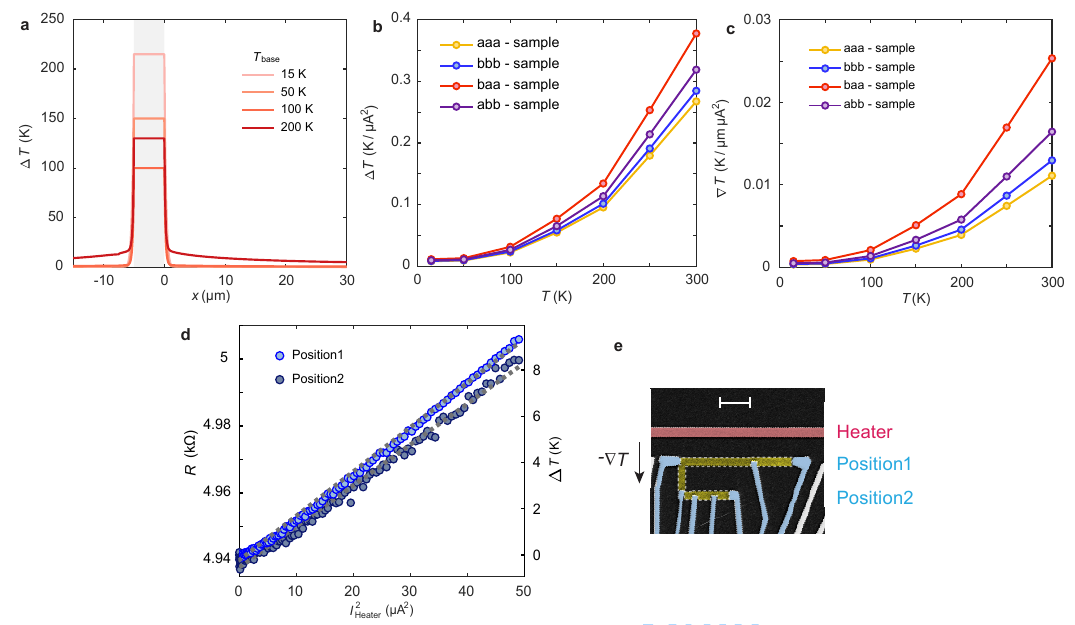}   
    }
    \caption{\textbf{a}, The simulated temperature profile on the surface of Si/SiO$_2$ substrates for various base temperatures corresponding to an applied heater current of 2 mA. The gray shaded range corresponds to the position of the Pt strip with a fixed temperature estimated from the temperature rise of the Pt electrode. \textbf{b}, The average temperature rise over the regions corresponding to the WTe$_2$ strip positions in the devices. \textbf{c}, The temperature gradient normal to the Pt heater electrodes for each of the device configuration used. \textbf{d}, Resistance of WTe$_2$ as a function of the square of the applied heater current for two different positions on the device, at a base temperature of 250 K. \textbf{e}, The false-coloured SEM image of the measured device and the corresponding voltage probe positions for panel \textbf{d} (Scale bar: $\SI{10} {\micro\meter}$). We further confirm that the temperature dependence of the resistance for this base temperature by performing an R-T calibration for both strips in order to extract the corresponding temperature difference derived from this calibration. This is shown as the right-hand $y$-axis in \textbf{d}.}
    \label{fig:simu_Si}
\end{figure}

Based on a similar procedure to that highlighted above, simulation of the temperature distribution was also performed for Si/SiO$_2$ (285 nm) substrate. The temperature profile as a function of the distance from the heater and the corresponding temperature rise and the temperature gradient for the WTe$_2$ flakes are shown in Figs.~\ref{fig:simu_Si}\textbf{a-c}. \\

We note that since the thermal conductivity of Si is several orders of magnitude larger than that of SiO$_2$ over the entire temperature range studied (see Table~\ref{tab:material}), Si/SiO$_2$ substrates act as much more efficient heat sink in comparison to fused silica substrates. As a result, the in-plane temperature gradient within the device is significantly (more than two order of magnitude) smaller across the flakes for devices on Si/SiO$_2$ substrates (Fig.~\ref{fig:simu_Si}\textbf{c}) in comparison to those on fused SiO$_x$ substrates (Fig.~\ref{fig:simu_fusedSiO2}\textbf{f}). Furthermore, the thermal conductivity of Si increases initially upon lowering the temperature, leading to a significantly smaller temperature gradient in WTe$_2$ which results in a smaller in-plane temperature gradient at lower temperatures.\\

To further verify that the simulation results are consistent with the temperature distribution in the device, we experimentally measured the temperature rise of WTe$_2$ in the measured devices. 
Fig.~\ref{fig:simu_Si}\textbf{d} shows the heater-current dependence of the resistance of WTe$_2$ for the sample used to quantify the anisotropic thermoelectric responses in WTe$_2$, measured at two different positions as illustrated in Fig.~\ref{fig:simu_Si}\textbf{e}.\\ 

We note that the rise in resistance corresponding to the rise in temperature follows proportionally to the square of the heater current at both positions, with slight differences in magnitude owing to their differing positions from the heater electrode. In this temperature regime, we further measure experimentally the relationship between the resistance and the base temperature to extract the temperature rise of the flake as a dunction of the heater current. The corresponding temperature difference between the two positions is shown on the same plot in Fig.~\ref{fig:simu_Si}\textbf{d}. From this, we ensure that the measured temperature rise and the differences are in good agreement with the simulation results.

\begin{table}[H]
\centering
\begin{threeparttable}
\begin{tabular}{|c|c|c|c|c|}
\hline
\textbf{Temperature (K)} & \textbf{amorphous SiO$_x$}$^{*}$ & \textbf{Si}$^{*}$ & \textbf{Au}\cite{ho1972thermal} & \textbf{WTe$_2$}\cite{qian2018anisotropic}  \\
\hline
10  & 0.1296 & 2339.2 & 3240 & - \\ \hline
50  & 0.3393 & 2662.8 & 421 & - \\ \hline
100 & 0.6721 & 885.31 & 327 & 21.0 \\ \hline
300 & 1.385 & 147.76 & 317 & 13.5 \\
\hline
\end{tabular}

\begin{tablenotes}[para]
\footnotesize
\centering
\item[*] From the COMSOL Material Library~\cite{COMSOL}.
\end{tablenotes}

\end{threeparttable}
\caption{Thermal conductivity [$\kappa$ (W\,m$^{-1}$\,K$^{-1}$)] of each material at representative temperatures}
\label{tab:material}
\end{table}

\subsection{The effect of the thermal conductivity of the materials under study} \label{Sec:Grad_OOP}

\begin{figure}[htb]
    \centering
   \makebox[\textwidth][c]{%
    \includegraphics[width=1.08\textwidth]{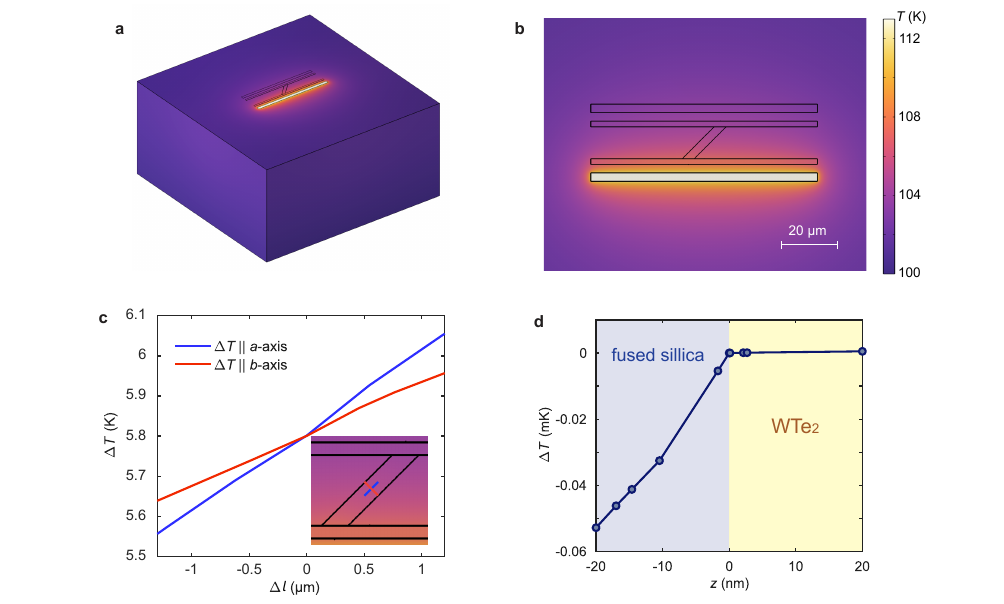}   
    }
    \caption{\textbf{a}, Three-dimensional simulation model and \textbf{b}, a magnified top-view image of the NLMT device on fused silica substrate. The resulting temperature distribution at a base temperature of 100 K is shown as a color map. \textbf{c}, Temperature profile along the $a$-axis and $b$-axis on the bottom surface of WTe$_2$ at for a bse temperature of 100 K. The line along which the profile is plotted is indicated in the schematics of the inset image. \textbf{d}, Temperature distribution normal to the device plane (along $z$-axis) at a base temperature of 100 K. The temperature at the interface between the substrate and WTe$_2$ (defined as $z = 0$) is taken as the reference, and the temperature difference ($\Delta$T = T(z) - T(z=0)) with respect this reference is shown. The $x,y$ positions correspond to the point where the two line profiles intersect in panel \textbf{c}.} 
    \label{fig:simu_WTe2}
\end{figure}

In addition to the COMSOL simulations performed on the simplified cross-sectional model, we additionally performed simulations of a three dimensional model. Figures ~\ref{fig:simu_WTe2}\textbf{a} and \textbf{b}  show the simulation model and a representative result for the NLMT device on a fused silica substrate. The device geometry is identical to that of the actual WTe$_2$ device used for NLMT measurements, aside from the substrate size.\\

The anisotropy of the temperature gradient within the flake in the NLMT configuration was evaluated using this model. Although an anisotropy in the thermal conductivity of WTe$_2$ has not been experimentally reported, theoretical predictions~\cite{liu2016first} and the measured anisotropy in the electrical conductivity suggest that the thermal conductivity of WTe$_2$ is also anisotropic. To account for this anisotropy, we estimate the ratios of the thermal conductivity along each crystallographic axis through the Wiedemann–Franz law, based on the measured anisotropy in the electrical conductivity (Fig.~\ref{fig_res_anisotropy}). These estimated values were then used in the simulations.\\

Figure~\ref{fig:simu_WTe2}\textbf{c} shows the temperature profiles extracted along the two different line cuts within the WTe$_2$ flake, taken along the crystallographic $a$- and $b$-axes. From these profiles, we find that the temperature gradient perpendicular to the strip ($\nabla_b T$) is approximately 39\% smaller than that along the strip ($\nabla_a T$).
If we do not take the thermal conductivity of the flake into account, we emphasize that the temperature gradients along the $a$- and $b$-axes would be expected to be identical, since the heater strip is oriented at 45° with respect to the voltage detection channel, resulting in a geometrically symmetric configuration. However, our results indicate that the higher thermal conductivity of WTe$_2$, compared to that of the fused silica substrate, leads to more efficient heat propagation along the WTe$_2$ strip, which in this configuration is oriented along the crystallographic-\textit{a} axis. In addition, the thermal conductivity along the crystallographic-\textit{a} axis is expected to be higher than that along the $b$-axis, further enhancing the anisotropic temperature gradient in the devices. With the approximation of assuming an isotropic thermal conductivity, we find that the difference in the product $\nabla_a T .\nabla_b T$ in comparison to the case including the anisotropic thermal conductivity of WTe$_2$ to be less than 5\%. \\

The temperature gradient normal to the $a$-$b$ plane, oriented in the out-of-plane direction was estimated using this model (Fig.~\ref{fig:simu_WTe2}\bf{d}\normalfont). Owing to the large thermal conductivity of WTe$_2$ compared to that of amorphous SiO$_x$, we note that the temperature gradient along the out-of-plane direction is almost negligible for WTe$_2$. The temperature gradient within WTe$_2$ along the $z$-axis is estimated to be four orders of magnitude smaller than the temperature gradient within the $a$–$b$ plane. We thus validate the assumption throughout this work of neglecting the temperature gradient along the $z$-axis and attributing the observed responses to an in-plane oriented temperature gradient.\\

For TaIrTe$_4$, although fewer studies on thermal conductivity have been reported compared to WTe$_2$, the value reported at room temperature for TaIrTe$_4$ is 14.4 W/mK along $a$-axis ~\cite{zhu2025anisotropic}, which is comparable to that in WTe$_2$ and thus a similar analysis and evaluation can be extended to TaIrTe$_4$ as well.



\subsection{Influence of contacts on the temperature distribution}
\begin{figure}[H]
    \centering
   \makebox[\textwidth][c]{%
    \includegraphics[width=1.05\textwidth]{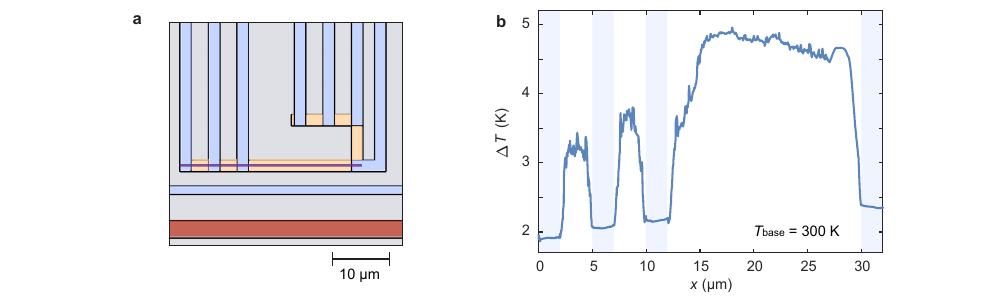}   
    }
    \caption{\textbf{a}, Magnified top view of the three-dimensional simulation model and, \textbf{b}, the calculated temperature profile along the WTe$_2$ strip. WTe$_2$, the Au contacts, and the Pt heater are shown in light orange, light blue, and red, respectively. The position of the line profile is indicated by the purple line in panel \textbf{a}, and the regions corresponding to the Au contacts are shaded in light blue in panel \textbf{b}.
    } 
    \label{fig:contact_temp}
\end{figure}

We note that the contact electrodes to the flakes have a significant influence on the temperature distribution within the device owing to the contacts acting as heat sinks. Figure~\ref{fig:contact_temp} shows the three-dimensional simulation model that reproduces the device geometry used for the anisotropic thermoelectric response measurements, along with the corresponding simulation results. From the line cut of the temperature distribution along the WTe$_2$ strip, a clear reduction in the local temperature is observed in the regions wherein the Au electrodes are placed, with temperatures approximately 1-2 K lower than the surrounding region that does not have the Au leads (Fig.~\ref{fig:contact_temp}\textbf{b}). This behavior arises from the much higher thermal conductivity of the electrodes compared to both the substrate and WTe$_2$ (see Table~\ref{tab:material}), causing them to act as effective heat sinks. As a consequence, we emphasize that if the contact electrodes are placed asymmetrically with respect to the flake in the lateral direction, the temperature distribution within the sample becomes correspondingly asymmetric. This leads to a parasitic longitudinal temperature gradient, as illustrated in Fig.~\ref{fig:contact_temp}\textbf{b}.\\

Based on these results, the devices used for nonlinear thermoelectric measurements presented in the main text were designed such that the contact electrodes are as symmetric as possible with respect to the flakes,to minimize the effect of parasitic thermal gradients (as shown in Figs.~1\textbf{b}–\textbf{d} in the main text).

\newpage

\bibliographystyle{apsrev4-2}
\bibliography{Reference_SI}

\end{document}